\documentclass[twocolumn,iop]{emulateapj}
\usepackage{apjfonts}
\usepackage{color}

\usepackage{amsmath,graphicx,longtable,hyperref}
\usepackage{natbib,threeparttable,rotating}
\usepackage{ulem}

\newcommand{\SiII}{\makebox{[Si{\sc\,II}]\,}}

\newcommand{\etal}{et al.}
\newcommand{\hbeta}{H{$\beta$}}
\newcommand{\halpha}{H{$\alpha$}}
\newcommand{\lya}{Ly\,$\alpha$}
\newcommand{\CIV}{C\,{\sevenrm IV}}

\newcommand{\SiIV}{Si\,{\sevenrm IV}}
\newcommand{\CIII}{C\,{\sevenrm III]}}

\newcommand{\AlIII}{Al\,{\sevenrm III}}

\newcommand{\SiIII}{Si\,{\sevenrm III]}}

\def\FeII{Fe\,{\sc ii}}
\def\MgII{Mg\,{\sc ii}}
\def\HeII{He\,{\sc ii}}
\def\HeIIopt{He\,{\sc ii}\,$\lambda$4687}
\def\HeIIuv{He\,{\sc ii}\,$\lambda$1640}
\def\CaII{Ca\,{\sc ii}}
\def\CaIIa{Ca\,{\sc ii}\,$\lambda$3934}

\newcommand{\OII}{[O{\sevenrm\,II}]}

\newcommand{\NeV}{[Ne\,{\sevenrm\,V}]}

\newcommand{\OIIIb}{[O{\sevenrm\,III}]\,$\lambda$5007}

\newcommand{\NIIb}{[N\,{\sevenrm\,II}]\,$\lambda$6584}
\newcommand{\NIIab}{[N\,{\sevenrm\,II}]\,$\lambda\lambda$6548,6584}

\newcommand{\SIIa}{[S\,{\sevenrm\,II}]\,$\lambda$6717}

\newcommand{\SIIab}{[S\,{\sevenrm\,II}]\,$\lambda\lambda$6717,6731}

\newcommand{\bracket}[1]{\left\langle#1\right\rangle}
   \font\sevenrm=cmr7 scaled 1000
\newcommand{\comments}[1]{}

\def\kms{{\rm km\,s^{-1}}}

\def\ergs{${\rm erg\,s^{-1}}$}

\begin{document}

\title{The Sloan Digital Sky Survey Reverberation Mapping Project: Sample Characterization}

\author{Yue Shen$^{1,2,*}$, Patrick B. Hall$^3$, Keith Horne$^4$, Guangtun Zhu$^5$, Ian McGreer$^6$, Torben Simm$^7$, Jonathan R.~Trump$^8$, Karen Kinemuchi$^9$, W.~N. Brandt$^{10,11,12}$, Paul J. Green$^{13}$, C.~J. Grier$^{10,6}$, Hengxiao Guo$^{2}$, Luis C.~Ho$^{14,15}$, Yasaman Homayouni$^{8}$, Linhua Jiang$^{14,15}$, Jennifer I-Hsiu Li$^{1}$, Eric Morganson$^{2}$, Patrick Petitjean$^{16}$, Gordon T.~Richards$^{17}$, Donald P. Schneider$^{10,11}$, D.~A. Starkey$^{1,4}$, Shu Wang$^{1,14}$, Ken Chambers$^{18}$, Nick Kaiser$^{18}$, Rolf-Peter Kudritzki$^{18}$, Eugene Magnier$^{18}$, Christopher Waters$^{18}$} 

\altaffiltext{1}{Department of Astronomy, University of Illinois at Urbana-Champaign, Urbana, IL 61801, USA; shenyue@illinois.edu}
\altaffiltext{2}{National Center for Supercomputing Applications, University of Illinois at Urbana-Champaign, Urbana, IL 61801, USA}
\altaffiltext{3}{Department of Physics and Astronomy, York University, Toronto, ON M3J 1P3, Canada}
\altaffiltext{4}{SUPA Physics and Astronomy, University of St Andrews, Fife, KY16 9SS, Scotland, UK}
\altaffiltext{5}{Department of Physics and Astronomy, Johns Hopkins University, 3400 N. Charles Street, Baltimore, MD 21218, USA}
\altaffiltext{6}{Steward Observatory, University of Arizona, 933 North Cherry Avenue, Tucson, AZ 85721-0065, USA}
\altaffiltext{7}{Max-Planck Institute for Extraterrestrial Physics, Giessenbachstrasse, Postfach 1312, 85741, Garching, Germany}
\altaffiltext{8}{University of Connecticut, Department of Physics, 2152 Hillside Road, Unit 3046, Storrs, CT 06269-3046, USA}
\altaffiltext{9}{Apache Point Observatory and New Mexico State University, P.O. Box 59, Sunspot, NM, 88349-0059, USA}
\altaffiltext{10}{Department of Astronomy \& Astrophysics, The Pennsylvania State University, University Park, PA, 16802, USA}
\altaffiltext{11}{Institute for Gravitation and the Cosmos, The Pennsylvania State University, University Park, PA 16802, USA}
\altaffiltext{12}{Department of Physics, 104 Davey Lab, The Pennsylvania State University, University Park, PA 16802, USA }
\altaffiltext{13}{Harvard-Smithsonian Center for Astrophysics, 60 Garden Street, Cambridge, MA 02138, USA}
\altaffiltext{14}{Department of Astronomy, School of Physics, Peking University, Beijing 100871, China}
\altaffiltext{15}{Kavli Institute for Astronomy and Astrophysics, Peking University, Beijing 100871, China}
\altaffiltext{16}{Institut d'Astrophysique de Paris, Sorbonnes Universit\'es, CNRS, 98bis, Boulevard Arago, 75014 Paris, France}
\altaffiltext{17}{Department of Physics, Drexel University, 3141 Chestnut Street, Philadelphia, PA 19104, USA}
\altaffiltext{18}{Institute for Astronomy, University of Hawaii at Manoa, Honolulu, HI 96822, USA}
\altaffiltext{$^*$}{Alfred P. Sloan Research Fellow.}

\shorttitle{SDSS-RM: Sample Characterization}
\shortauthors{Shen \etal}

\begin{abstract}

We present a detailed characterization of the 849 broad-line quasars from the Sloan Digital Sky Survey Reverberation Mapping (SDSS-RM) project. Our quasar sample covers a redshift range of $0.1<z<4.5$ and is flux-limited to $i_{\rm PSF}<21.7$ without any other cuts on quasar properties. The main sample characterization includes: 1) spectral measurements of the continuum and broad emission lines for individual objects from the coadded first-season spectroscopy in 2014; 2) identification of broad and narrow absorption lines in the spectra; 3) optical variability properties for continuum and broad lines from multi-epoch spectroscopy. We provide improved systemic redshift estimates for all quasars, and demonstrate the effects of signal-to-noise ratio on the spectral measurements. We compile measured properties for all 849 quasars along with supplemental multi-wavelength data for subsets of our sample from other surveys. The SDSS-RM sample probes a diverse range in quasar properties, and shows well detected continuum and broad-line variability for many objects from first-season monitoring data. The compiled properties serve as the benchmark for follow-up work based on SDSS-RM data. The spectral fitting tools are made public along with this work.
\keywords{
black hole physics -- galaxies: active -- line: profiles -- quasars: general -- surveys
}
\end{abstract}

\section{Introduction}\label{sec:intro}

The Sloan Digital Sky Survey Reverberation Mapping (SDSS-RM) project is a dedicated multi-object RM campaign that simultaneously monitors 849 quasars covering a wide redshift and luminosity range with moderate-cadence imaging and spectroscopy \citep[for a technical overview of the SDSS-RM project, see][]{Shen_etal_2015a}. Started in 2014, SDSS-RM will continue through 2020 to build a spectroscopic time baseline of seven years, and a photometric time baseline of a decade when combining dedicated SDSS-RM imaging (2014-2020) with earlier (2010-2013) imaging from the Pan-STARRS 1 \citep[PS1,][]{Kaiser_etal_2010} survey. 

The primary science goal of SDSS-RM is to measure RM lags between continuum flux and broad emission line flux \citep[e.g.,][]{Blandford_McKee_1982,Peterson_1993}, which provide an estimate of the typical size of the broad-line region (BLR). Combining the measured RM lag and the width of the broad emission lines, one can estimate a ``virial'' mass of the central black hole (BH) assuming the BLR is virialized. RM is the primary technique to measure BH masses in active galaxies, and it anchors the so-called ``single-epoch'' BH mass estimators \citep[for a  review, see, e.g.,][]{Shen_2013}, where the latter have been extensively used in the field to estimate quasar BH masses at all redshifts with single-epoch spectroscopy. SDSS-RM aims to extend previous RM studies that were limited to low-$z$ and mostly low-luminosity AGN to both higher redshifts and a more representative sample by targeting a flux-limited ($i<21.7$) quasar sample at $0.1<z<4.5$ without any cuts on quasar properties. The multiplex capability of SDSS-RM greatly improves the efficiency of RM, and our 2014 data set already led to lag detections that expand the redshift-luminosity range of RM measurements \citep[e.g.,][]{Shen_etal_2016a,Li_etal_2017,Grier_etal_2017}.

In addition to lag measurements, the multi-epoch images and spectroscopy from SDSS-RM enable a diverse range of applications from host galaxy properties to accretion disk properties \citep[e.g.,][]{Shen_etal_2015b,Shen_etal_2016b,Matsuoka_etal_2015,Grier_etal_2015,Sun_etal_2015,Sun_etal_2018,Denney_etal_2016a,Denney_etal_2016b,Li_etal_2017,Yue_etal_2018,Homayouni_etal_2018}. The SDSS-RM field also has extensive multi-wavelength coverage from previous surveys and our dedicated follow-up programs. The SDSS-RM sample is a representative sample of quasars over a broad range of redshifts and physical parameters. As such, it is a highly valuable dataset for quasar science, and a detailed characterization of the sample properties will be beneficial to many follow-up studies of the SDSS-RM sample.   

In this work we present a detailed characterization and compilation of quasar properties for the SDSS-RM sample. In \S\ref{sec:data} we describe the sample and the data; in \S\ref{sec:spec} we detail our spectral characterization of the sample, and in \S\ref{sec:var} we describe our optical variability characterization; we describe our compiled catalogs in \S\ref{sec:cat} and summarize in \S\ref{sec:sum}. Additional information regarding our spectral fitting code is provided in the appendix. Throughout this paper we adopt a flat $\Lambda$CDM cosmology with $\Omega_\Lambda=0.7$ and $h=0.7$.

\section{Data}\label{sec:data}

\subsection{Sample Overview}


The SDSS-RM field is a single 7 ${\rm deg^2}$ field that coincides with the PS1 Medium Deep Field \citep[MDF;][]{Tonry_etal_2012b} MD07 (R.A. J2000=213.704, decl. J2000=+53.083), which lies within the CFHT-LS W3 field\footnote{http://www.cfht.hawaii.edu/Science/CFHTLS/}.

The SDSS-RM sample contains 849 broad-line quasars over a redshift range of $0.1<z<4.5$. Most of the SDSS-RM quasars were previously spectroscopically confirmed quasars in SDSS I-III. In particular, the SDSS-III BOSS survey \citep{Dawson_etal_2013} targeted and spectroscopically confirmed most of these quasars. As detailed in \citet{Ross_etal_2012}, the BOSS quasar survey combines optical color selection with a variety of ancillary target selection criteria to increase the total number of quasars. All objects classified as point-like and having magnitudes of $g<22$ or $r < 21.85$ are passed to the quasar target selection code. A small number of quasars were added to the sample that were missed from SDSS; these objects were targeted with optical variability (55 quasars) and multi-wavelength selection (7 quasars) with follow-up spectroscopic confirmation \citep[see details in sec 3.1 of][]{Shen_etal_2015a}. Table \ref{tab:format} provides the original target flags used in SDSS to select these quasars, which follow the same definitions as in the SDSS I-III surveys, and a flag ``OTHER\_TARGET'' to indicate additional target selection. The SDSS-RM quasar sample is intended to be a nearly complete, flux-limited sample of unobscured broad-line quasars, without any other cuts on multi-wavelength properties, emission line properties, or variability properties. With additional target selection using optical variability or mid-infrared photometry from the Wide-field Infrared Survey Explorer (WISE) survey \citep[][]{Wright_etal_2010}, the SDSS-RM sample is more complete in optical color space contaminated by stars \citep[e.g.,][]{Schmidt_etal_2010,Morganson_etal_2015} and less biased against dust-reddened type 1 quasars \citep[e.g.,][]{Assef_etal_2013} than optical-only color selection. {Nevertheless, obscured type 2 quasars without broad emission lines are apparently excluded from our sample, and some heavily dust-reddened quasars with peculiar colors may still be missing from our sample.} 

\begin{figure}
\centering
    \includegraphics[width=0.48\textwidth]{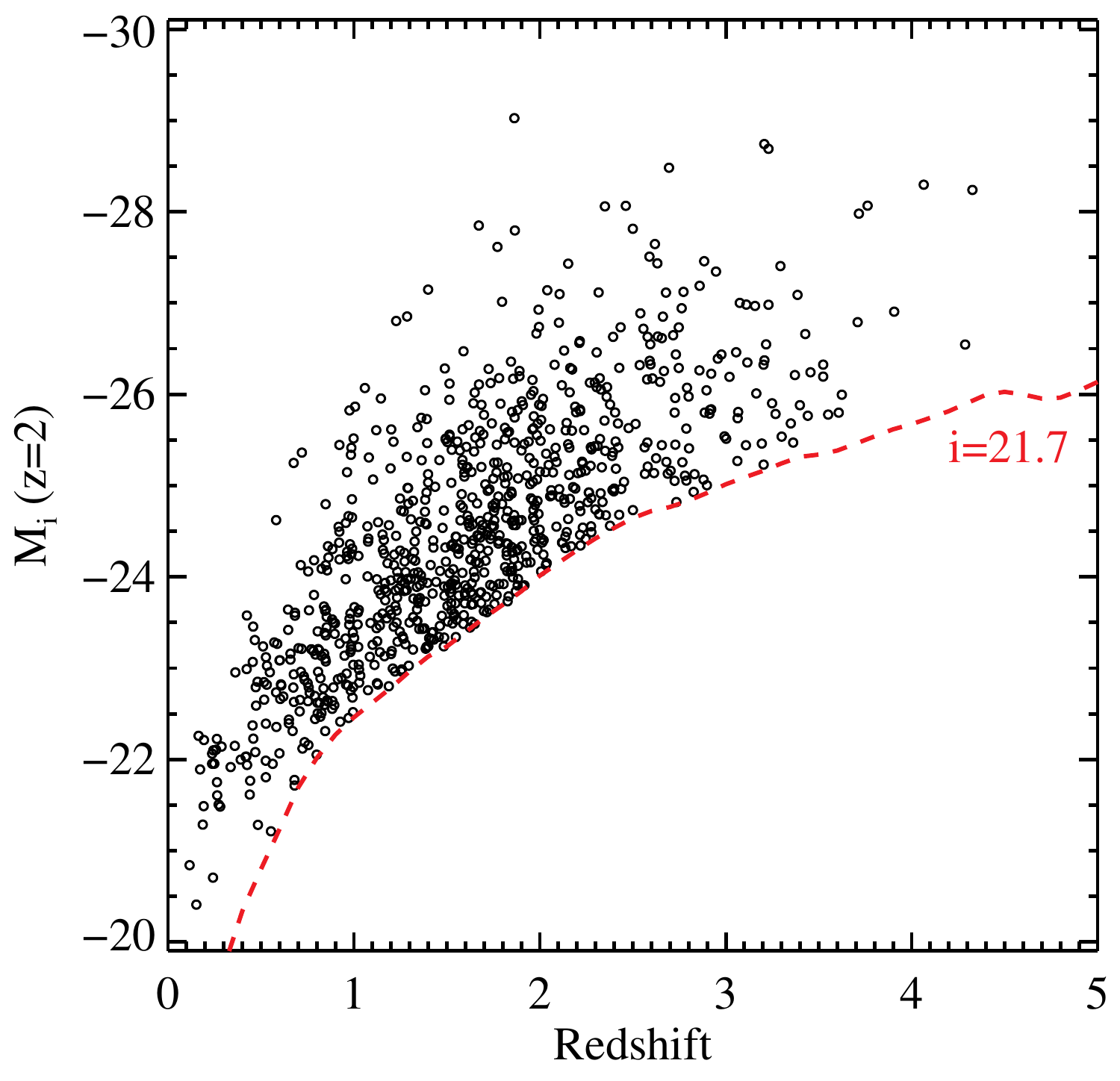}
    \caption{Distribution of the SDSS-RM quasar sample in the redshift-luminosity space. We use the emission-line flux corrected, absolute $i$ band magnitude normalized at $z=2$ (equivalent to the rest-frame 2500\,\AA\ luminosity) as the luminosity indicator, following \citet{Richards_etal_2006a}. The formal flux limit of the SDSS-RM sample is $i=21.7$ (PSF magnitude, uncorrected for Galactic extinction).}
    \label{fig:Lz}
\end{figure}

Fig.\ \ref{fig:Lz} displays the redshift-luminosity distribution of the SDSS-RM sample. Following \citet{Richards_etal_2006a}, we use the emission line corrected, absolute $i$ band magnitude normalized at $z=2$ (equivalent to the rest-frame 2500\,\AA\ luminosity), $M_i(z=2)$, as the continuum luminosity indicator. {The conversion between $M_i(z=2)$ and the conventional magnitude $M_i(z=0)$ is \citep[][their eqn. 1]{Richards_etal_2006a}: $M_i(z=0)=M_i(z=2)+0.596$}. The formal flux limit of the sample is $i<21.7$ (PSF magnitude, uncorrected for Galactic extinction). There is a deficit of targets near the flux limit at low redshift, due to the incomplete target selection in this regime as the main quasar target selection in BOSS was biased against extended sources \citep{Ross_etal_2012}. Compared to earlier RM samples that predominately focus on low-redshift and low-luminosity AGN, the SDSS-RM sample covers a much broader and contiguous range in redshift and luminosity, spanning almost all of cosmic time and range from {Seyferts ($M_i(z=2)\gtrsim -22.5$ or $L_{\rm bol}\lesssim 10^{45}\,{\rm erg\,s^{-1}}$) to luminous quasars}. 

\subsection{SDSS-RM Optical Imaging and Spectroscopy}


The primary spectroscopy has been obtained with the SDSS telescope \citep[][]{Gunn_etal_2006} and the BOSS spectrographs \citep[][]{Smee_etal_2013}. The first-season SDSS-RM spectroscopic data were taken from 2014 January to July in SDSS-III \citep[][]{Eisenstein_etal_2011} and consist of a total of 32 epochs with an average cadence of $\sim 4$ days; each epoch had a typical exposure time of 2\,hr. The SDSS-RM program has continued in SDSS-IV \citep[][]{Blanton_etal_2017}, with $\sim 12$ epochs per year (2 per month) with a nominal exposure time of 1\,hr each during 2015-2017, and $\sim 6$ epochs per year (monthly cadence) during 2018-2020. As of July 2018 we have obtained a total of 78 spectroscopic epochs and a spectroscopic baseline of 5 years (2014-2018). The wavelength coverage of BOSS spectroscopy is $\sim 3650-10400$\,\AA, with a spectral resolution of $R\sim 2000$. The typical signal-to-noise ratio (S/N) per 69 $\kms$ pixel averaged over the $g$ band in a 2 hr exposure is $\sim 4.5$ at $g_{\rm psf}=21.2$, but could be lower for epochs observed with poor observing conditions. 

We obtained additional spectroscopic data with MMT/Hectospec in 2017 in order to test the feasibility of continuing the spectroscopic monitoring with other facilities, which covers most of the RM targets in our sample as well as ancillary science targets such as high-$z$ quasar candidates and variable stars. These data will be presented elsewhere. 

Supporting photometric observations in both the $g$ and $i$ bands have been obtained primarily with the Steward Observatory Bok 2.3m telescope on Kitt Peak and the 3.6m Canada-France-Hawaii Telescope (CFHT) on Mauna Kea, which roughly cover the same monitoring period as SDSS-RM spectroscopy with a cadence of $\sim 2$ days in 2014 and reduced cadences in successive years. Details of the photometric observations and the subsequent data processing will be presented by Kinemuchi et al. (2018). In addition, we also have early multi-band photometric light curves from PS1 during 2010-2013 with a typical cadence of several days. These early PS1 light curves substantially extend the photometric baseline of SDSS-RM, and are critical to measuring long lags (e.g., on multi-year timescales) when combined with the later SDSS-RM spectroscopy. The SDSS-RM field continues to be monitored with the PS1 system in 5 bands since 2016, albeit with reduced cadence ($\sim$ weekly to monthly per band).

By the end of the program in 2020, SDSS-RM will have a time baseline of more than a decade for imaging and seven years for spectroscopy.

\begin{table}
\caption{Line Fitting Parameters}\label{tab:linefit}
\centering
\scalebox{1.0}{
\begin{tabular}{cccc}
\hline\hline
Line Complex & Fitting Range [\AA] & Line &  $n_{\rm gauss}$ \\
(1) & (2) & (3) & (4)  \\
\hline
\halpha & 6400-6800 & broad \halpha & 3 \\
  & & narrow \halpha & 1 \\
  & & [NII]6549   & 1 \\
  & & [NII]6585   & 1 \\
  & & [SII]6718   & 1 \\
  & & [SII]6732   & 1 \\
\hbeta  & 4640-5100 & broad \hbeta & 3 \\
  & & narrow \hbeta & 1 \\
  & & [OIII]4959 core &  1 \\
  & & [OIII]5007 core & 1 \\
  & & [OIII]4959 wing & 1 \\
  & & [OIII]5007 wing & 1 \\
  & & broad HeII 4687 & 1 \\
  & & narrow HeII 4687 & 1 \\
\MgII & 2700-2900 & broad \MgII\ & 2 \\
  & & narrow \MgII\ & 1 \\
\CIII\ & 1700-1970 & \CIII & 2 \\
 & & \SiIII\,1892 & 1 \\
 & & \AlIII\,1857 & 1 \\
 & & \SiII\,1816 & 1 \\
 & & NIII 1750 & 1 \\
 & & NIV1718 & 1 \\
\CIV & 1500-1700 & \CIV & 3 \\
 & & broad HeII 1640 & 1 \\
 & & narrow HeII 1640 & 1 \\
 & & broad OIII 1663 & 1 \\
 & & narrow OIII 1663 & 1 \\
\SiIV & 1290-1450 & broad SiIV/OIV] & 2 \\
 & & CII 1335 & 1 \\
 & & OI 1304 & 1 \\
\lya & 1150-1290 & \lya\ & 3 \\
 & & NV 1240 & 1 \\
\hline
\hline\\
\end{tabular}
}
\begin{tablenotes}
      \small
      \item NOTE. --- Our emission line fits are performed in individual line complexes, where multiple lines are fit simultaneously. The last column lists the number of Gaussians used for each line. 
\end{tablenotes}
\end{table}

\subsection{Other Multi-Wavelength Data}

The SDSS-RM field is covered by several wide-area sky surveys such as the WISE survey \citep[][]{Wright_etal_2010} and the FIRST radio survey \citep[][]{White_etal_1997}. We compile the WISE photometry from the AllWISE release, as well as forced WISE photometry at the positions of SDSS sources \citep[unWISE,][]{Lang_etal_2016}. 

For the FIRST radio matches, we follow the procedures in \citet{Jiang_etal_2007} and \citet{Shen_etal_2011}: we match the SDSS-RM quasars with the FIRST source catalog (14 Dec, 2017 version) with a matching radius of 30\arcsec\ and estimate the radio loudness $R = f_{\rm 6 cm}/f_{2500}$, where $f_{\rm 6 cm}$ and $f_{2500}$ are the flux density ($f_\nu$) at rest-frame 6 cm and 2500\AA, respectively. For quasars with only one FIRST source within 30\arcsec\ we match them again to the FIRST catalog with a matching radius of 5\arcsec and classify the matched ones as core-dominant radio quasars. Quasars with multiple FIRST source matches within 30\arcsec\ are classified as lobe-dominated. The rest-frame 6 cm flux density is determined from the FIRST integrated flux density at 20 cm assuming a power-law slope of $\alpha_\nu = -0.5$; the rest-frame 2500\AA\ flux density is determined from our spectral fits. For lobe-dominated radio quasars, we use all the matched FIRST sources to compute the radio flux density. 

The SDSS-RM field also overlaps with several extragalactic multi-wavelength fields, notably the AEGIS field \citep{Davis_etal_2007}. We have collected public multi-wavelength data for a subset of SDSS-RM quasars. These data include the multi-wavelength data from \citet{Nandra_etal_2015} for 32 SDSS-RM quasars, and Spitzer IRAC and MIPS data from the Spitzer Enhanced Imaging Products (SEIP) source list for 176 SDSS-RM quasars. Since these data apply only to a small subset of the SDSS-RM sample, we provided these compiled data in separate ancillary table files and archived them on the SDSS-RM data server \footnote{ftp://quasar.astro.illinois.edu/public/sdssrm/paper\_data/Sample\_char/}. 

Finally, there are GALEX NUV light curves and coadded imaging data for about half of the SDSS-RM sample from the GALEX Time Domain Survey \citep{Gezari_etal_2013}, UKIRT near-IR imaging for the RM field, and X-ray data from various XMM-Newton and Chandra programs. In particular, we have obtained XMM-Newton imaging for the entire SDSS-RM field and are compiling X-ray properties for the sample. We will compile these future datasets when they become available and distribute through the SDSS-RM data server. 

\section{Spectral Characterization}\label{sec:spec}

To measure the continuum and line fluxes from the spectra, we use a spectral fitting approach similar to our earlier work \citep[e.g.,][]{Shen_etal_2008,Shen_etal_2011,Shen_Liu_2012} but with a number of modifications, which we describe in detail below.

In this work we use the coadded 2014 spectra (32 epochs in total). The 32 epochs were coadded using the SDSS-III spectroscopic pipeline {\tt idlspec2d}, to produce high S/N individual spectra for all 849 quasars in the SDSS-RM sample. The coadded spectra have identical format as other SDSS-III spectra, with a logarithmic wavelength binning of $10^{-4}$ in $\log_{10}\lambda$ and are stored in vacuum wavelengths.  

\begin{figure*}
\centering
    \includegraphics[height=0.9\textwidth,angle=-90]{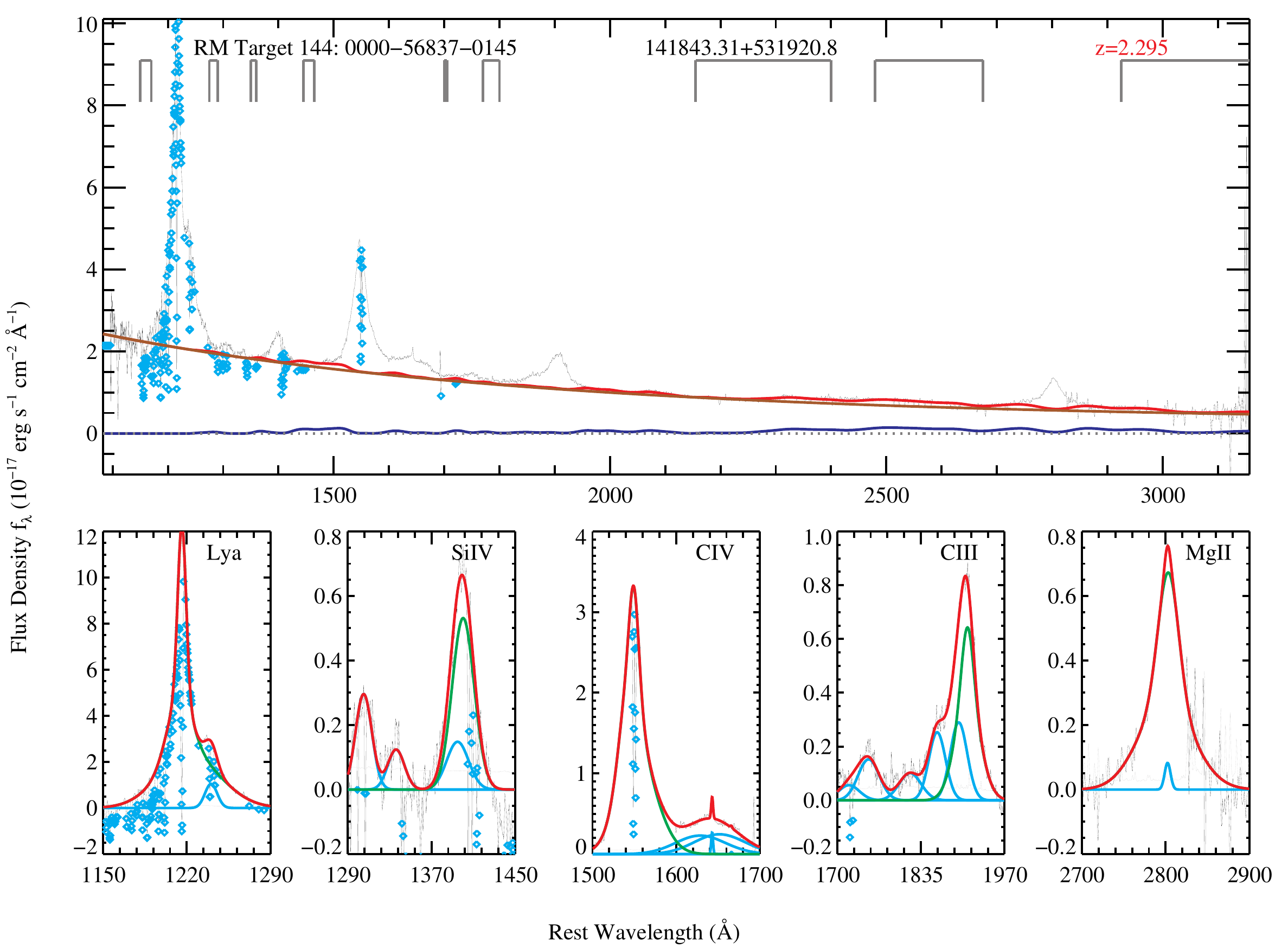}
    \caption{An example of our global spectral fitting approach. The top panel shows the continuum (brown) and \FeII\ (blue) model components; the red line is the sum of the two. The cyan diamonds are pixels masked as absorption or bad pixels. The gray brackets near the top of the panel indicate the windows used for the continuum+\FeII\ fit. The bottom panels present the emission line fits for five line complexes. }
    \label{fig:examp_fit}
\end{figure*}

\subsection{Spectral Fitting}

We fit the coadded spectra\footnote{The coadded spectra are distributed on the SDSS-RM data server, and are compiled in a single file (spPlate-0000-56837.fits) with the same format as other SDSS plates, where the fiber number corresponds to RMID+1 of the sample \citep[][]{Shen_etal_2015a}. } from the first-season (2014) observations with a continuum$+$emission line model. The high SNR of the coadded spectra for individual objects allows measurements of weak emission lines, and these measurements represent the average properties of the quasar during this monitoring period. The fitting was performed in the rest-frame of the quasar using the pipeline redshift, after correcting for Galactic reddening using the dust map in \citet{SFD} and the extinction curve from \citet{CCM}. 

The continuum fit was performed for all pixels in a set of relatively line-free (except for broad-band \FeII\ emission) windows over the entire spectrum of the quasar. The continuum model is described by a power-law plus a 3rd-order polynomial, where the additive (positive-definite) polynomial component is introduced to fit objects with peculiar (e.g., bending) continuum shapes likely caused by peculiar intrinsic dust reddening (see an example in Fig.\ \ref{fig:int_obj}, top left panel). In addition to the continuum, we fit the optical and UV \FeII\ emission using empirical templates from the literature \citep[e.g.,][]{Boroson_Green_1992, Vestergaard_Wilkes_2001,Tsuzuki_etal_2006,Salviander_etal_2007}. The continuum and the \FeII\ emission form a pseudo-continuum, which is subtracted from the spectrum to form a line-only spectrum for which we measure emission line properties. We do not include a Balmer continuum component in the fit because fitting such a feature requires sufficient wavelength coverage that is not available for most of our quasars; but our fitting code implements such a feature, which can be switched on and off. We also do not include a host galaxy component since such a component cannot be well constrained for most of our objects at $z\gtrsim 1$. Empirical corrections for host starlight for the low-$z$ subset of our quasars are discussed in \S\ref{sec:host}.

The continuum and \FeII\ emission models are given by the following equations: 

\begin{eqnarray}
f_{\rm pl}(\lambda) & = & a_0(\lambda/\lambda_0)^{a_1}\ , \\
f_{\rm poly}(\lambda) & = & \sum_{i=1}^3 b_{i}(\lambda - \lambda_0)^i\ , \\
f_{\rm FeII}(\lambda) & = & c_0F_{\rm FeII}(\lambda, c_1, c_2)  \ , \\
f_{\rm conti}(\lambda) & = & f_{\rm pl} + f_{\rm poly} + f_{\rm FeII}\ ,
\end{eqnarray}
where $\lambda_0=3000$\,\AA\ is the reference wavelength, and $a_i$, $b_i$, $c_i$ are the model parameters; specifically $c_1$ and $c_2$ are the Gaussian broadening and wavelength shift parameters applied to the \FeII\ templates to match the data.  

We then fit the line spectrum in logarithmic wavelength space, i.e., the natural binning scheme of SDSS spectra. The fitting is performed on individual line complexes where multiple lines are close in wavelength. Table \ref{tab:linefit} lists the detailed information of line complexes and the fitting parameters. In each line complex, we fit a set of Gaussians to individual lines, with constraints on their velocities and widths. We generally do not constrain the flux ratio of line doublets, but we fix the flux ratio\footnote{We fix the flux ratio for \SIIab\ because the lines are weak; we fix the flux ratio for \NIIab\ to reduce ambiguity of line decomposition under the \halpha\ profile.} of the \SIIab\ doublet to be 1 and the flux ratio of the \NIIab\ doublet to be $f_{6584}/f_{6548}=3$. For most permitted lines, we attempt to decompose the narrow-line component and the broad-line component. However, for weak lines and lines that lack a distinctive division between broad and narrow lines (such as \CIV) we do not attempt such a decomposition. Fig.\ \ref{fig:examp_fit} shows an example of our model fit for quality assessment (QA); the full set of QA plots are available on the SDSS-RM data server. 

Many high-$z$ quasars have UV absorption lines imprinted on the spectrum, which may bias the continuum and emission line fits. To remedy the effect of absorption lines, in the fit we mask pixels that are 5$\sigma$ below the 20-pixel boxcar smoothed spectrum. In addition, we perform one iteration after rejecting pixels that fall 3$\sigma$ below the previous fit. The combination of these two absorption rejection criteria is a good recipe for correcting the effects of narrow absorption lines, and somewhat improves the fits for broad absorption line quasars.

We measure continuum luminosities and emission line properties (line peak, FWHM, equivalent width, etc.) from the model fits to the spectrum. These spectral properties are compiled in the main catalog described in Table \ref{tab:format}. In addition to the primary broad and narrow emission lines, we include measurements for the narrow \OII\,$\lambda$3728 and \NeV\,$\lambda$3426 lines and the stellar absorption line \CaII\,$\lambda$3934 (K). As discussed in \citet{Shen_etal_2016b}, these relatively isolated lines often suffer from imperfect global continuum subtraction. Hence we refit these lines with a local continuum model and report the measurements in Table \ref{tab:format}.

To estimate the measurement uncertainties of the spectral properties, we use a Monte Carlo approach: we add to the original spectral flux at each pixel a random Gaussian variate with zero mean and $\sigma$ given by the reported error at that pixel and repeat the fitting procedure on the mock spectrum; we create 50 trials and estimate the measurement uncertainty of each spectral quantity (e.g., continuum flux, line FWHM, etc.) as the semi-amplitude of the range enclosing the 16th and 84th percentiles of the distribution from the trials. {Adding flux perturbations to the original spectrum instead of the model spectrum preserves details in the spectral features (such as absorption lines) that are not captured or well-fit by our model. On the other hand, the original spectrum is already a perturbed version of the noise-free true spectrum, hence the mock spectra are slightly noisier than the original spectrum, and therefore our approach will produce more conservative measurement errors in the spectral quantities. }

Additional information about the global fitting code is provided in Appendix \ref{sec:app2}.

\subsection{SNR Dependence}

The high SNR of the coadded spectra also allows investigation of the SNR dependence of our spectral measurements. Similar analyses were performed to evaluate the robustness of the measured line peaks in \citet{Shen_etal_2016b} and \CIV\ line widths in \citet{Denney_etal_2016a}; here we extend this exercise to a more complete list of lines and spectral quantities. We degrade the original high-SNR spectra by inflating the original flux errors by a constant scaling factor and perturbing the spectrum using the new errors. We then perform the same fitting procedure on the degraded spectra, and compare the results with those based on the original high-SNR spectra. This approach isolates the effect of SNR from that of the intrinsic variability of the quasar on the spectral measurements, but still explores the full range of spectral diversity.

\begin{figure*}
\centering
    \includegraphics[height=0.8\textwidth,angle=-90]{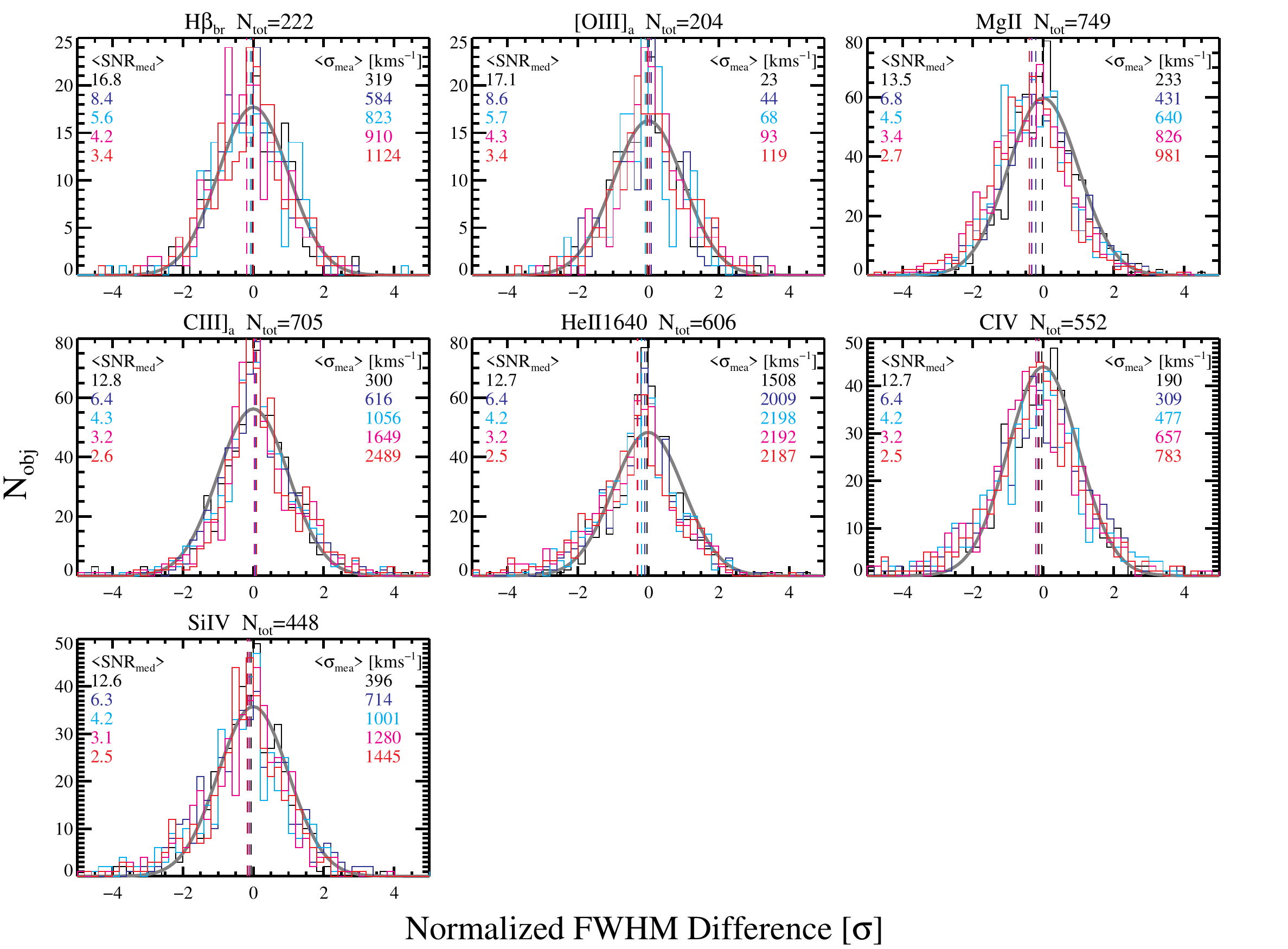}
    \includegraphics[height=0.8\textwidth,angle=-90]{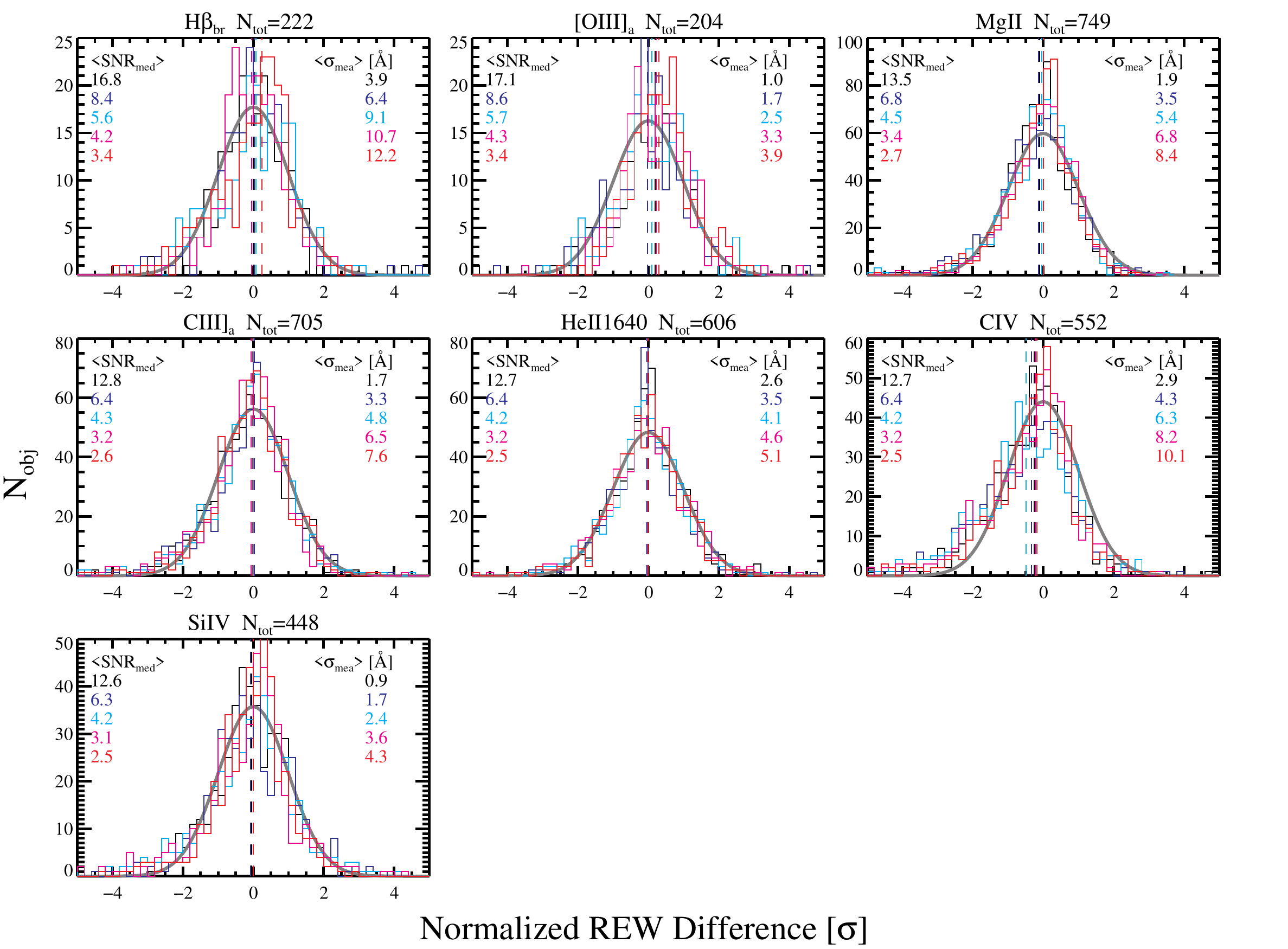}
    \caption{Distribution of the difference in the measured line FWHM (upper panel) and rest EW (lower panel) between the original high SNR spectra and the degraded spectra, normalized by the measurement uncertainty (the quadrature sum of the measurement uncertainties in both measurements). Each panel shows results for a specific line, with the total number of objects listed at the top. Different colors represent different degradations of the original SNR by a factor of 2, 4, 6, 8, and 10. The sample median of the median SNR across the spectral range (${\rm \bracket{SNR_{med}}}$) and the sample median measurement uncertainty (${\rm \bracket{\sigma_{mea}}}$) in the degraded spectra are marked in each panel. The median of the distribution is indicated by the vertical dashed line. The gray solid line is a Gaussian with zero mean and unity dispersion, normalized to have the same area as the observed distribution. For all lines and all cases of SNR degradation, there is negligible offset in the distribution, indicating that the measurement of line peak is unbiased when the continuum SNR is decreased to as low as $\sim 3$ per SDSS pixel. In addition, the agreement between the observed distributions and the unity Gaussian suggests that our estimated measurement uncertainties from the Monte Carlo approach are reasonable.}
    \label{fig:sn_test}
\end{figure*}

We have already demonstrated in \citet{Shen_etal_2016b} that the peak wavelengths of the lines can be measured reasonably well even if the median SNR of the spectrum is as low as $\sim 3$ per SDSS pixel, with negligible biases \citep[e.g., fig.\ 3 of][]{Shen_etal_2016b}. Fig.\ \ref{fig:sn_test} shows the difference (normalized by measurement uncertainties) between measurements from the degraded spectra and from the high SNR coadded spectra for typical narrow and broad lines of quasars, for the line FWHMs (upper panel) and rest-frame equivalent widths (REW, lower panel). As with the case of line peaks, there is no systematic bias in the measurements when the spectral SNR is degraded to as low as $\sim 3$ per SDSS pixel, consistent with our earlier findings using general SDSS quasars \citep[][]{Shen_etal_2011}. This comparison also demonstrates that our measurement errors estimated from the Monte Carlo approach are {reasonable (or even more conservative)} because they yield normalized distributions consistent with the expected Gaussian distribution with unity dispersion.  

It is also interesting to note that the FWHM of the \CIV\ line, which often displays asymmetric and blueshifted profiles, can be measured robustly with the multi-Gaussian function even at SNR of a few with negligible bias. This result differs from the case where a Gauss-Hermite function \citep[e.g.,][]{vanderMarel_Franx_1993} is used instead to fit the \CIV\ line, where a systematic bias with respect to SNR is observed \citep[e.g.,][]{Denney_etal_2016a}. As discussed in that paper, this behavior is due to the fact that the Gauss-Hermite model is more flexible than the multi-Gaussian model and tends to over-fit the line when the SNR is low, leading to a systematic bias in FWHM measurement. On the other hand, we adopted a 3-Gaussian model for the \CIV\ line in this work, as opposed to the 2-Gaussian model used for the comparison in \citet{Denney_etal_2016a}. For high SNR spectra, the 3-Gaussian model reproduces the \CIV\ line profile better than a 2-Gaussian model and just as well as the Gauss-Hermite model, but does not have a systematic bias as the SNR decreases as for the Gauss-Hermite model \citep[][]{Denney_etal_2016a}.  Based on these results, we conclude that the multi-Gaussian model is a better function to use for the \CIV\ line in general than the Gauss-Hermite model, particularly for survey-quality spectra. 

While we have focused on the general quasar population and concluded that SNR degradation does not lead to significant biases in the spectral measurements, it is possible that certain individual objects with peculiar line profiles are more sensitive to SNR degradation, which of course also depends on the flexibility of the fitting model.  

\subsection{Host-Galaxy Properties}\label{sec:host}

For low-redshift ($z\lesssim 1$) objects in the SDSS-RM sample, there could be significant host contamination in the continuum. We have used a spectral decomposition technique described in \citet{Vandenberk_etal_2006} and \citet{ShenJ_etal_2008} to decompose the spectrum into quasar and host-galaxy components. \citet{Shen_etal_2015b} provided details on the application of this spectral decomposition to the SDSS-RM quasar sample, and presented successful decomposition and measurements of the host stellar velocity dispersion in $\sim 100$ SDSS-RM quasars at $z\lesssim 1.1$. Consistent results were obtained with independent spectral decomposition in \citet{Matsuoka_etal_2015} and with imaging decomposition in \citet{Yue_etal_2018} for the same SDSS-RM sample. For convenience we have compiled the host fraction at $5100$\,\AA\ and stellar velocity dispersion measurements from \citet{Shen_etal_2015b} in the main catalog presented here. 

%
%
%

\subsection{Improved Redshifts}\label{sec:z}

With the measured line peaks from our spectral fits, we improve the systemic redshift estimation of SDSS-RM quasars following the recipes in \citet{Shen_etal_2016b}. In short, \citet{Shen_etal_2016b} recommended a ranked list of lines as redshift indicators using the measured line peak, calibrated to be consistent with those derived from the stellar absorption lines on average. We adopt the line-based redshift with the smallest combined systematic and measurement uncertainty as the systemic redshift. 

\begin{figure}
\centering
    \includegraphics[height=0.48\textwidth,angle=-90]{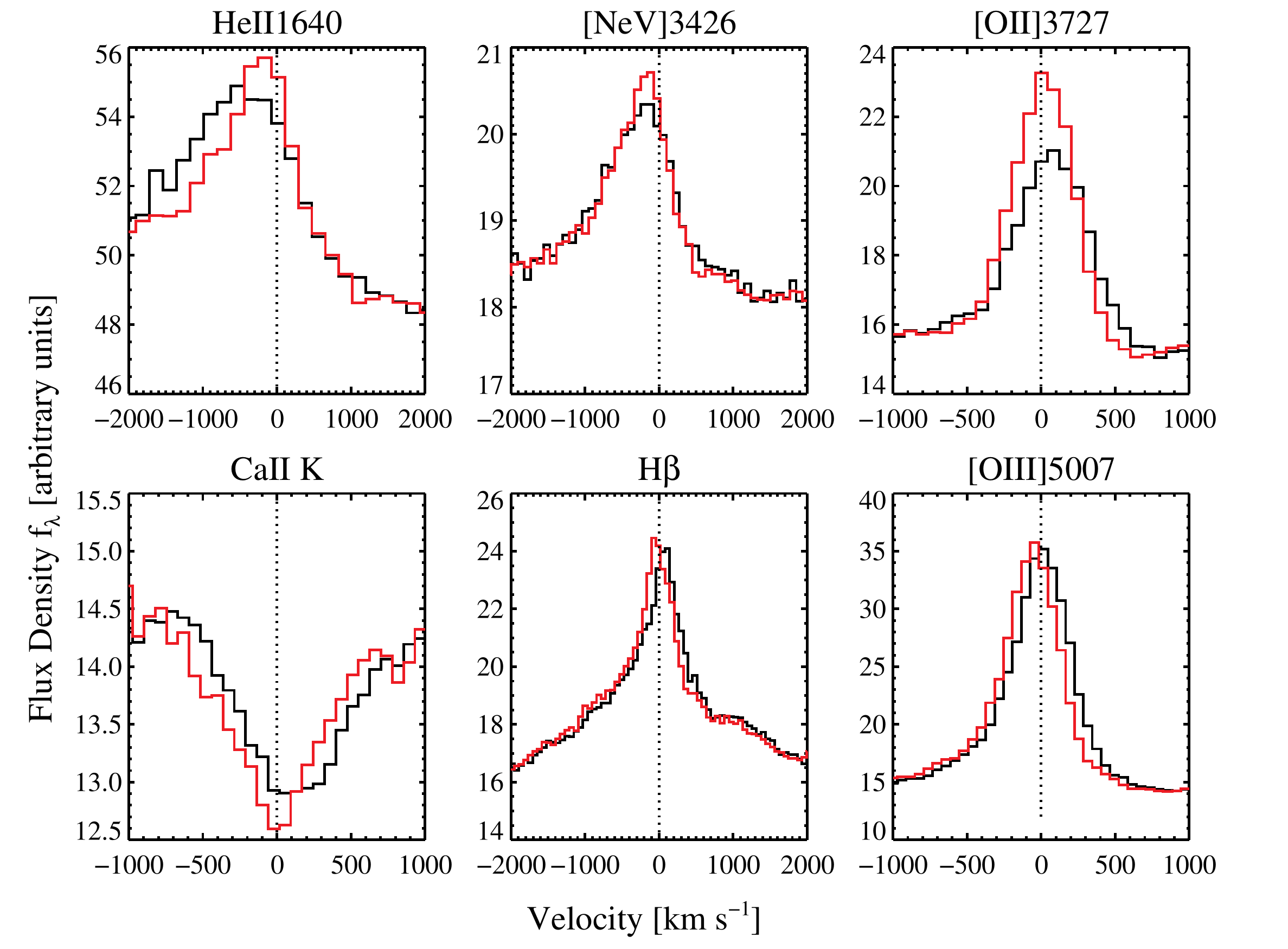}
     \caption{Median composite spectra around several emission/stellar absorption lines of the SDSS-RM sample, generated using the pipeline redshifts (black lines) and the improved systemic redshift (red lines). The systemic redshifts make the line features in the composite spectrum sharper or closer to the expected wavelengths. }
    \label{fig:zsys_comp}
\end{figure}

Fig.\ \ref{fig:zsys_comp} presents the comparison of the median composite spectra generated by coadding all SDSS-RM quasars with the pipeline redshifts and the improved systemic redshifts, following the methodology of generating composite spectrum in \citet{Vandenberk_etal_2001}. The lines in the composite spectrum with the improved redshifts are significantly sharper, and better aligned with the \CaIIa\ velocity, than those in the composite spectrum with the pipeline redshifts, indicating that our revised redshifts are more accurate than the pipeline redshifts. Most quasar emission lines, on average, have an intrinsic offset velocity from the systemic velocity based on stellar absorption lines, as determined in \citet{Shen_etal_2016b}. These average velocity offsets are reflected in the composite spectrum shown in Fig.\ \ref{fig:zsys_comp}.

\subsection{Absorption Line Systems}\label{sec:abs}

A significant fraction of high-$z$ quasars in the SDSS-RM sample show broad and/or narrow absorption lines in their spectra. We have identified these absorption line systems with independent approaches from our spectral fitting described above. These absorption line systems can be used to study the time variability of absorption using spectra from individual epochs as well as detailed abundance of the narrow absorbers with the coadded high-SNR spectra. Some initial results on the time variability of broad absorption troughs were reported in \citet{Grier_etal_2015} and Hemler et al. (submitted). 

\subsubsection{BALQSOs}

We identify 95 quasars with broad absorption line (BAL) troughs based on the 2014 spectra.

BAL and mini-BAL quasars
\citep[e.g.,][]{Trump_etal_2006,Gibson_etal_2009,Paris_etal_2017} were identified by their absorption troughs.
Troughs were visually identified by inspecting coadded spectra from the first year of observations (2014).
No limits were placed on the velocity of the absorption relative to
the quasar's systemic redshift.
BALs were searched for in \ion{C}{4}, \ion{Al}{3}, \ion{Mg}{2},
\ion{Fe}{2}, and \ion{Fe}{3}.

A maximum trough velocity width for each BAL quasar was estimated using
the widest trough in the 2014 coadded spectrum (whenever possible).
Velocity limits were set where the trough reached 90\% of the estimated
unabsorbed continuum.  That raw width was corrected to a true velocity
width of the absorbing gas by subtracting the velocity separation of
the doublet involved (\CIV\ or \MgII).

We adopt the convention that BAL quasars have troughs $> 2000$ km s$^{-1}$
wide and that mini-BAL quasars
\citep[e.g.,][]{Rodriguez_etal_2011}
have troughs $500 - 2000$ km s$^{-1}$ wide.
No troughs $<500$ km s$^{-1}$ wide were selected for our sample.
Neither do we include apparent mini-BAL troughs which are collections of
narrow \CIV\ absorbers clustered in velocity space (as revealed by
narrow absorption in other ions such as \ion{N}{5} or \ion{Si}{4}).
Therefore, from the perspective of automated identification of troughs
in a certain range of velocity widths, our sample is incomplete.
From the perspective of studying broad intrinsic absorption,
our sample should be highly complete: we rejected half a dozen borderline
mini-BAL candidates. 

We classify these BALQSOs into three classes: 1) HiBAL (72 quasars): quasars with \CIV\ BAL toughs only; 2) LoBAL (21 quasars): quasars with both \CIV\ and \MgII/\AlIII\ BALs; 3) FeLoBAL (2 quasars): LoBAL quasars with additional Fe BAL troughs. {Some HiBAL quasars may have LoBALs that were shifted out of the SDSS spectral coverage. }

We provide the list of 95 BAL quasars and notes on individual objects in a CSV file accompanying this paper, which is also distributed  in the SDSS-RM data server. 

\subsubsection{Narrow Absorbers}

We have identified narrow absorption lines (NALs) imprinted on the SDSS-RM quasar spectra following the approach described in detail in \citet{Zhu_Menard_2013}. The search for NALs was limited to 10 absorption systems per sight-line at different absorber redshifts, since the vast majority of SDSS-RM quasars have fewer NAL systems. To reduce the impact of broad absorption lines in high-ionization lines such as \CIV, which will make the identification of NALs difficult, we focus the search on \MgII\ absorption-line systems and associated lines (e.g., \FeII). Given the much higher S/N in the coadded SDSS-RM spectra than regular SDSS spectra, we were able to identify weaker absorption-line systems than those reported in \citet{Zhu_Menard_2013}.

For each NAL system identified, we measure the absorption EW and dispersion of a list of line species using Gaussian fits \citep[e.g.,][]{Zhu_Menard_2013}. Our main catalog compilation in Table \ref{tab:format} provides the redshifts of the identified NAL systems, and detailed measurements of these absorbers are provided in supplemental data files, with the same format as the NAL catalog presented in \citet{Zhu_Menard_2013}. 



\subsection{Example objects of special interest}

\begin{figure*}
\centering
    \includegraphics[width=0.48\textwidth]{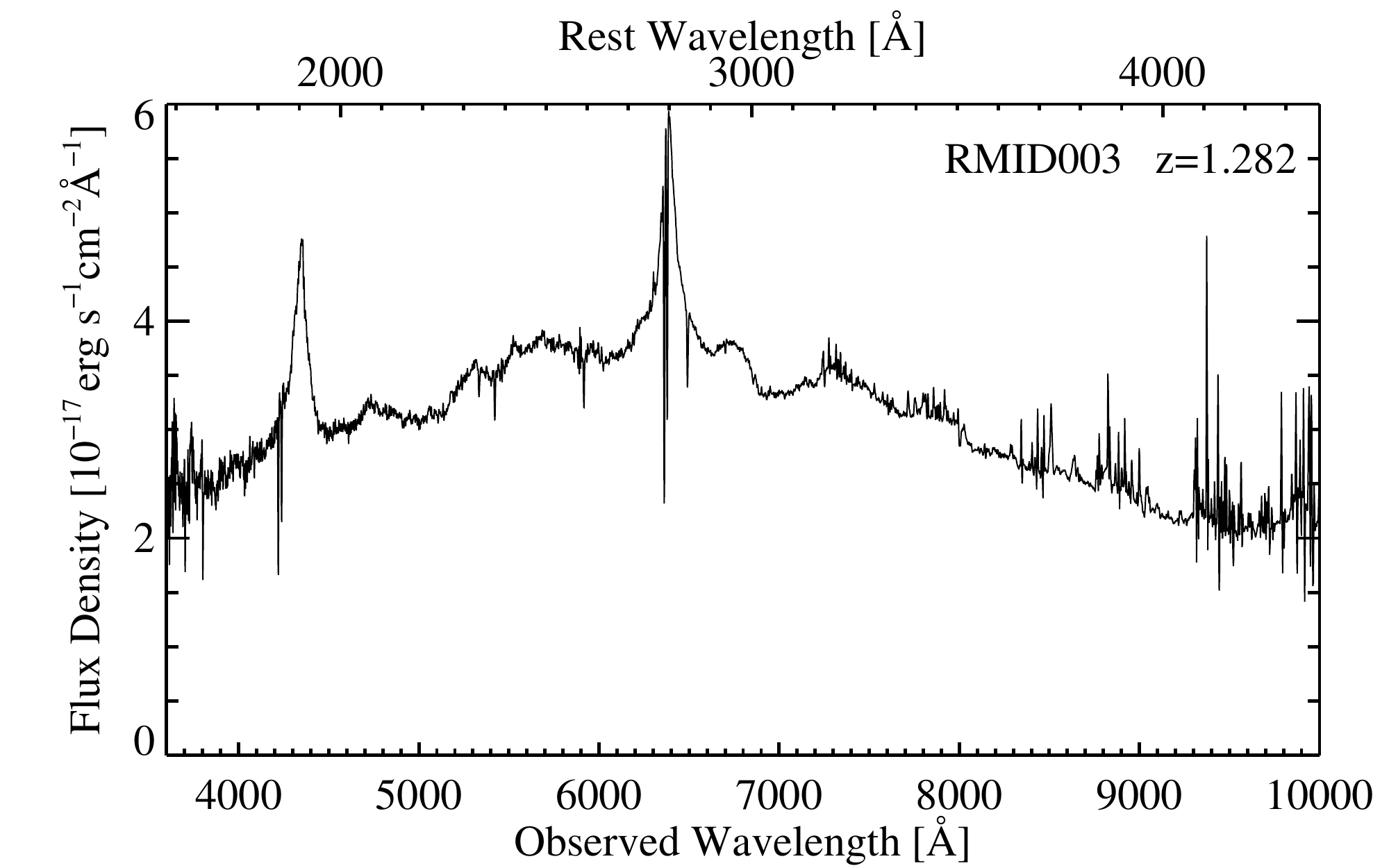}
    \includegraphics[width=0.48\textwidth]{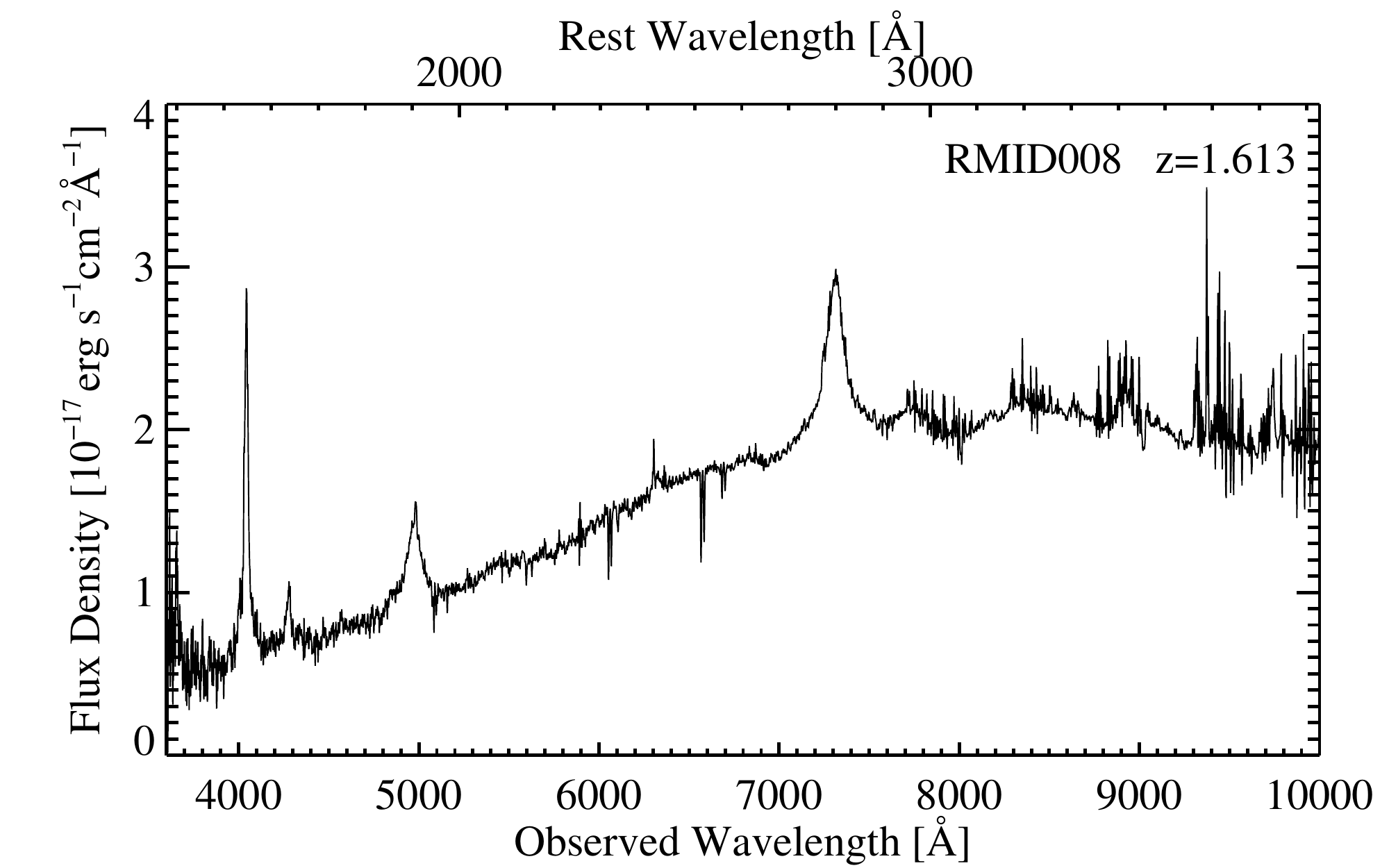}
    \includegraphics[width=0.48\textwidth]{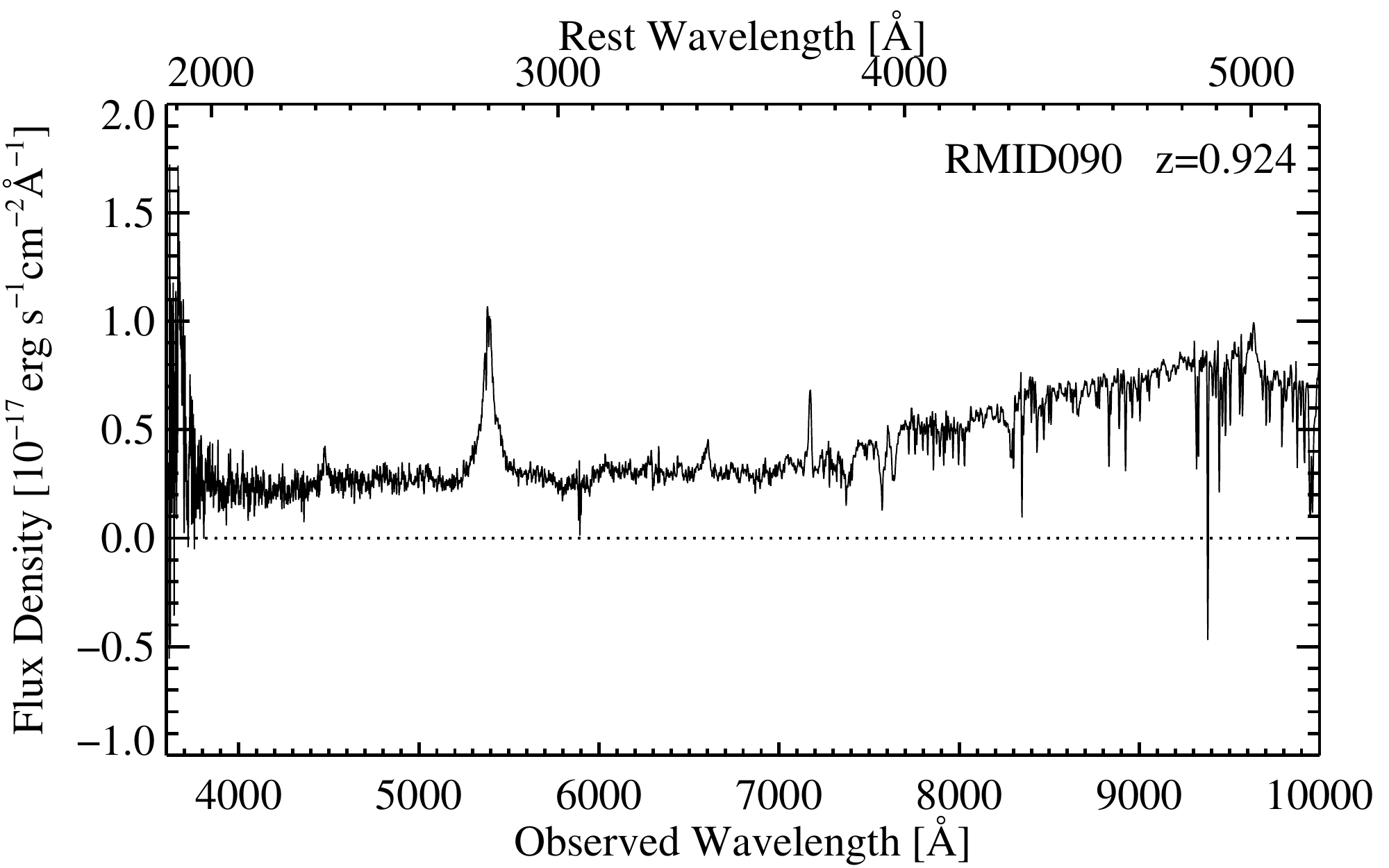}
    \includegraphics[width=0.48\textwidth]{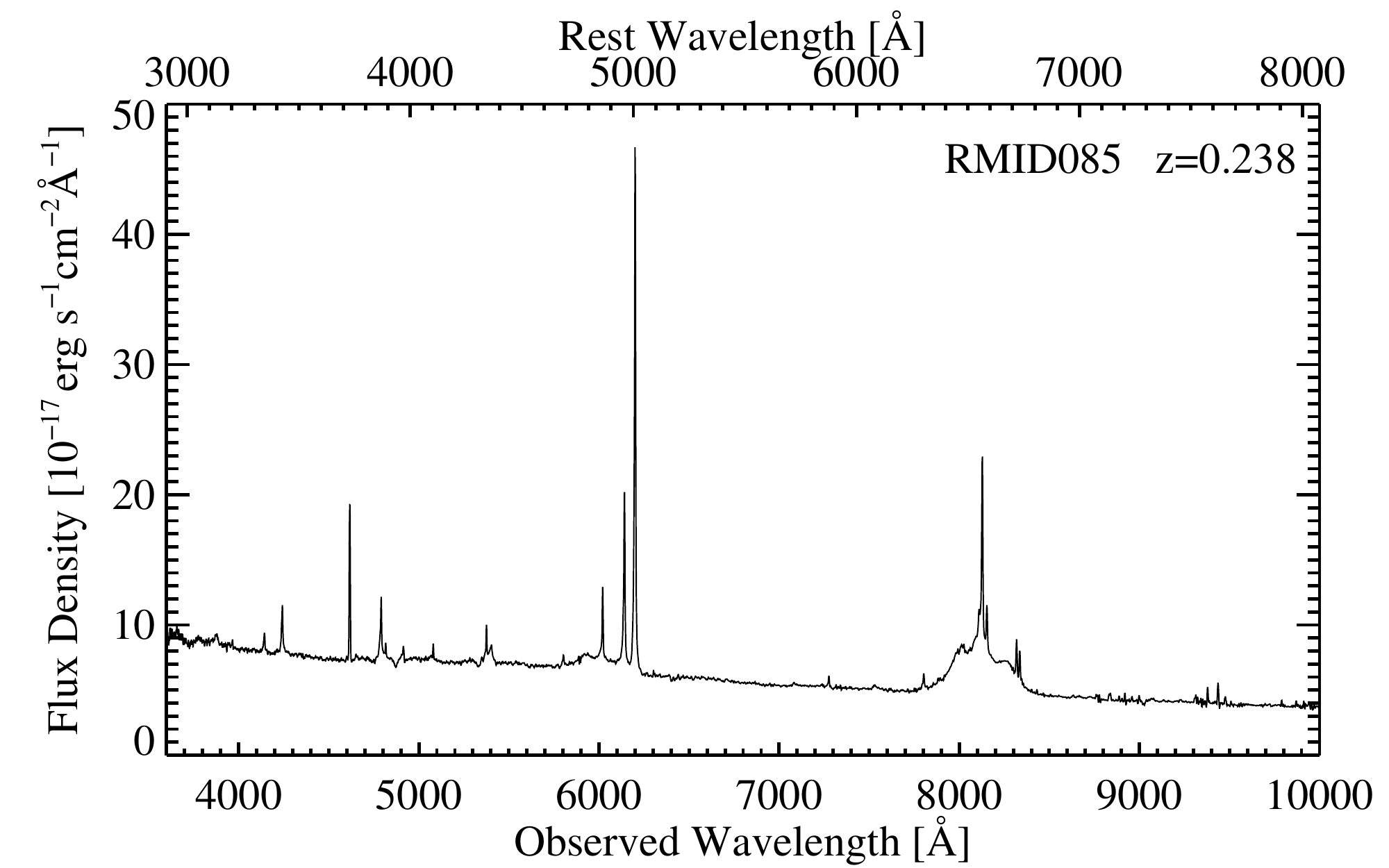}
    \includegraphics[width=0.48\textwidth]{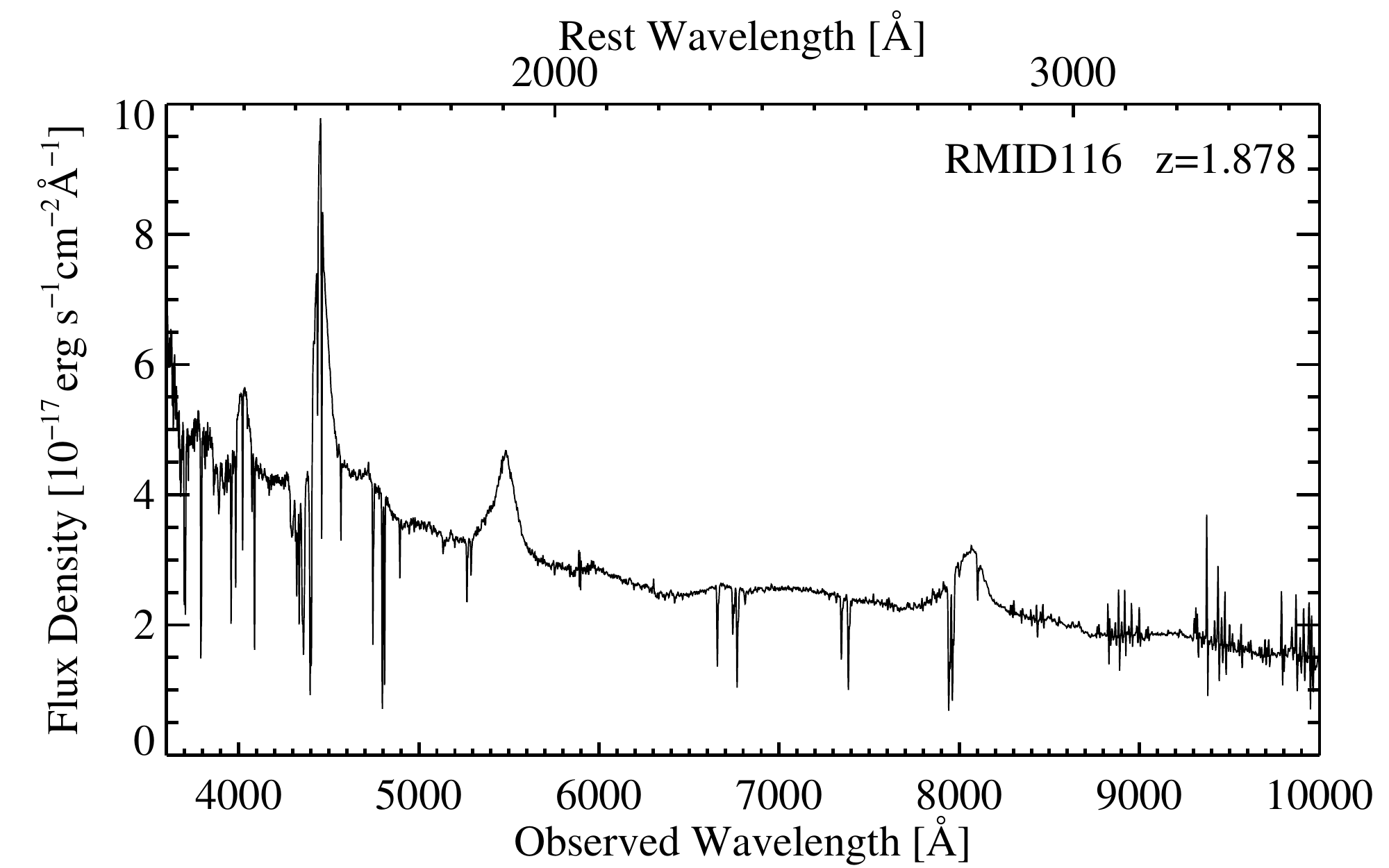}
    \includegraphics[width=0.48\textwidth]{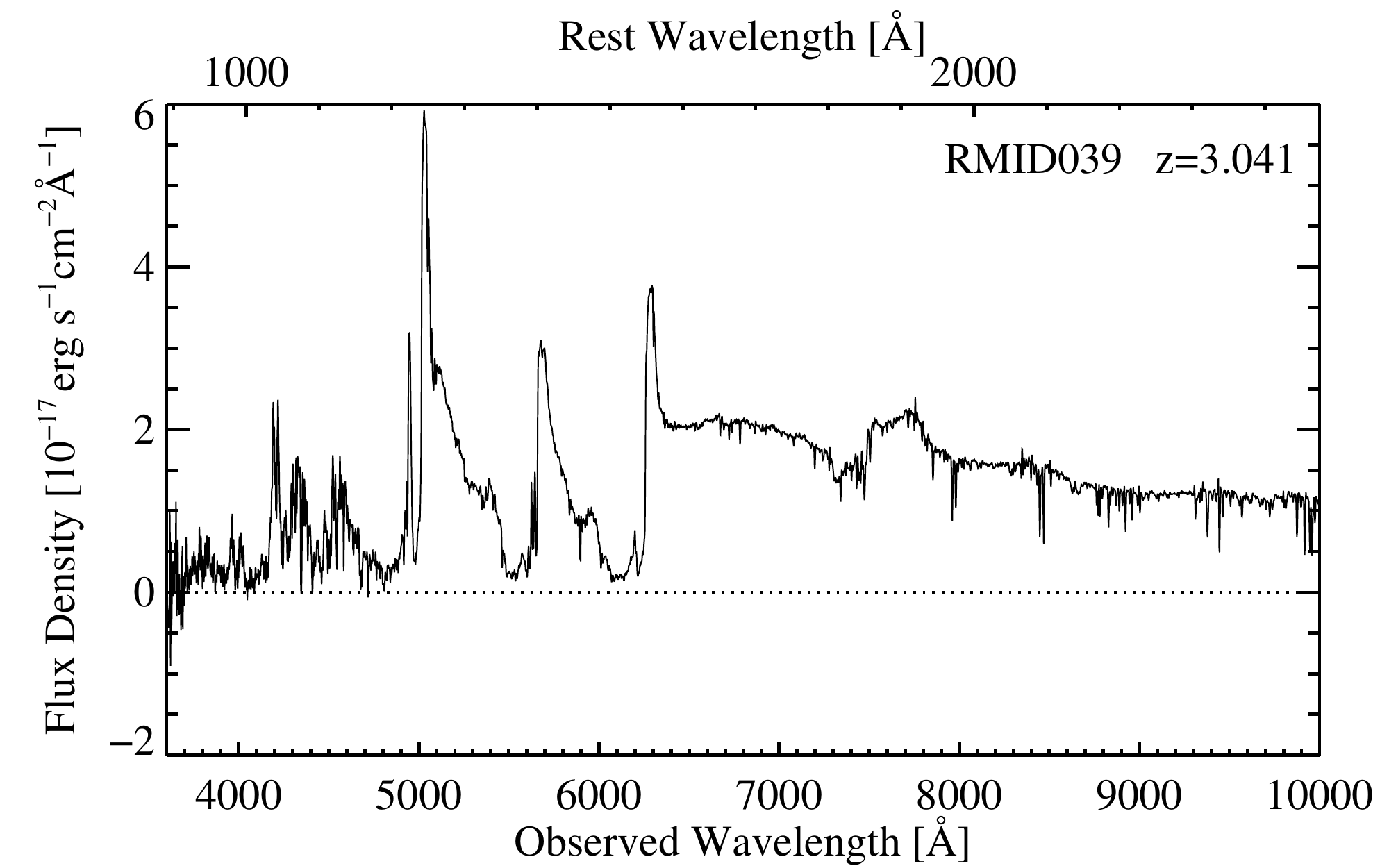}
    \includegraphics[width=0.48\textwidth]{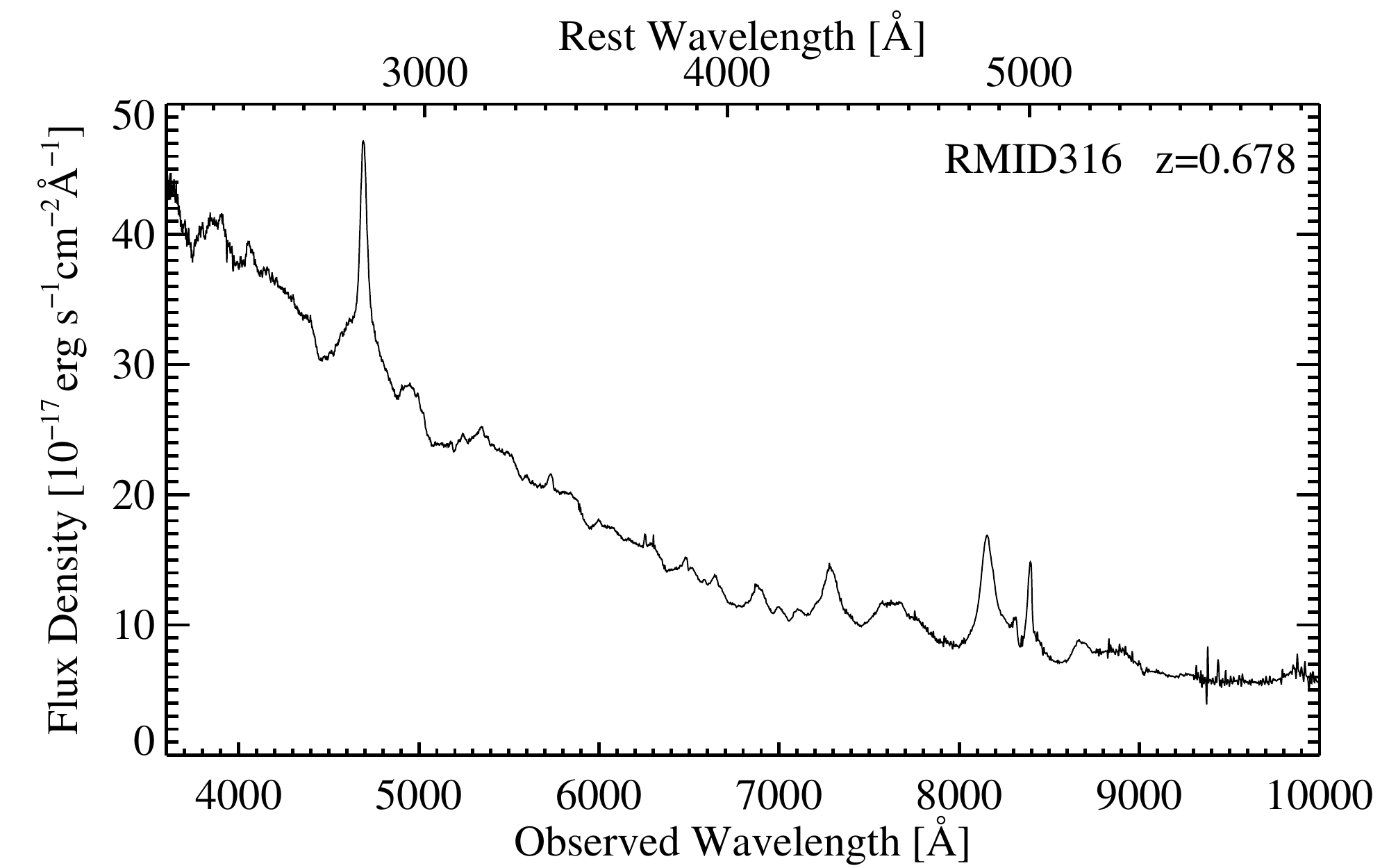}
    \includegraphics[width=0.48\textwidth]{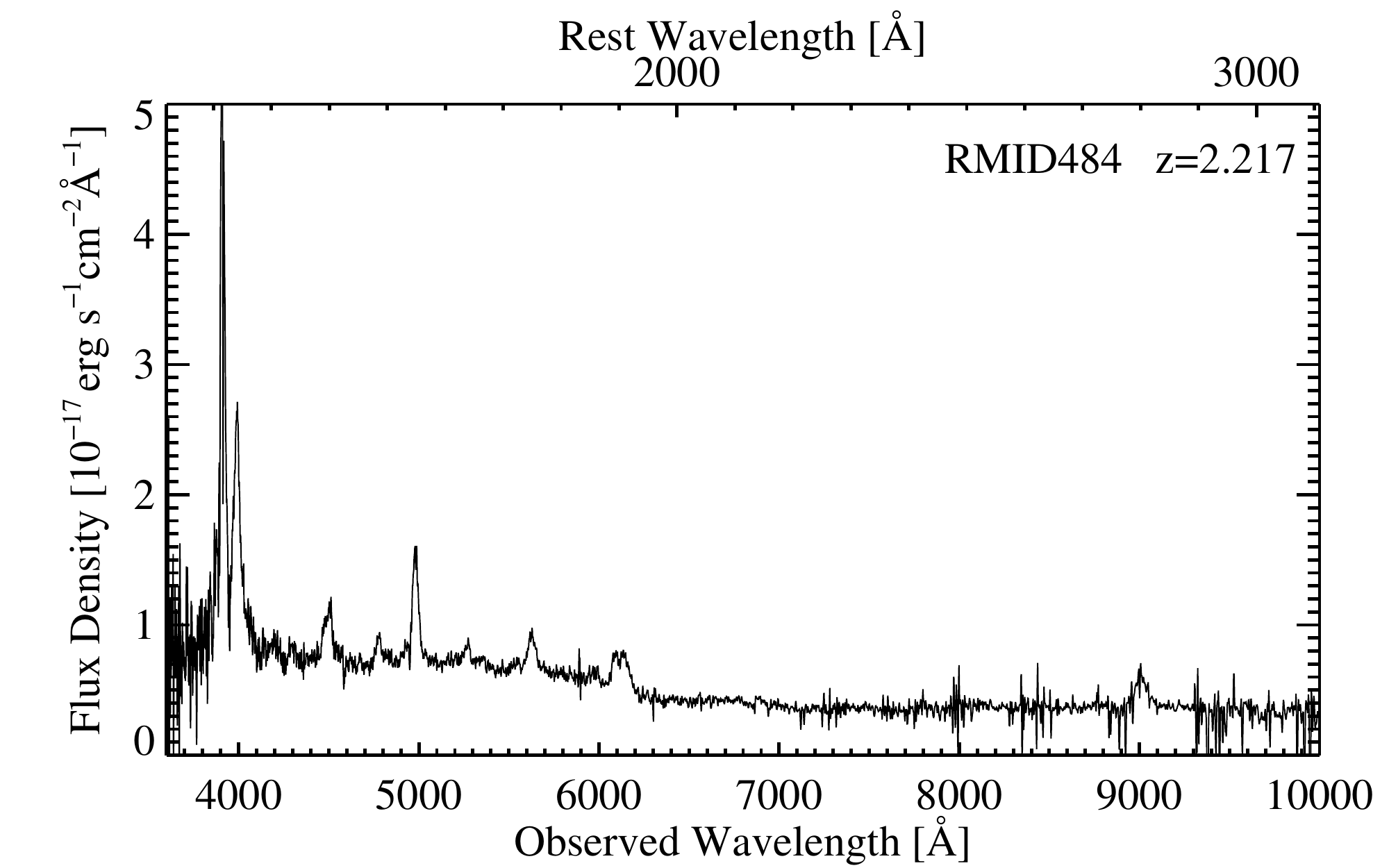}
     \caption{Objects in the SDSS-RM sample that are of particular interest. First row: two quasars with peculiar continuum shapes likely caused by peculiar intrinsic reddening; second row: a quasar with strong stellar absorption lines in rest-frame optical and broad rest-frame UV lines (left), and a quasar showing double-peaked broad Balmer lines (dubbed ``disk emitters''; right); third row: a quasar showing many narrow absorption lines (left), and a broad absorption line quasar (right); fourth row: an \FeII-rich quasar (left), and a nitrogen-rich quasar \citep[right, e.g.,][]{Jiang_etal_2008b,Liu_etal_2018a}.}
    \label{fig:int_obj}
\end{figure*}

The high SNR coadded spectra and the time-resolved individual spectra offer a wide range of applications to understand the physics of quasars and to study interesting individual systems. Figure \ref{fig:int_obj} presents several examples of quasars in the SDSS-RM sample that are of particular interest. These include peculiar dust reddened quasars, quasars with double-peaked broad emission lines \citep[dubbed ``disk emitters'',][]{Chen_etal_1989,Eracleous_Halpern_1994}, broad and narrow absorption line quasars, strong \FeII\ emitters, and quasars with unusually strong nitrogen emission \citep{Jiang_etal_2008}. 

%
%
%
%
%
%
%

\section{Optical Variability Characterization}\label{sec:var}

\begin{figure}
\centering
    \includegraphics[width=0.48\textwidth]{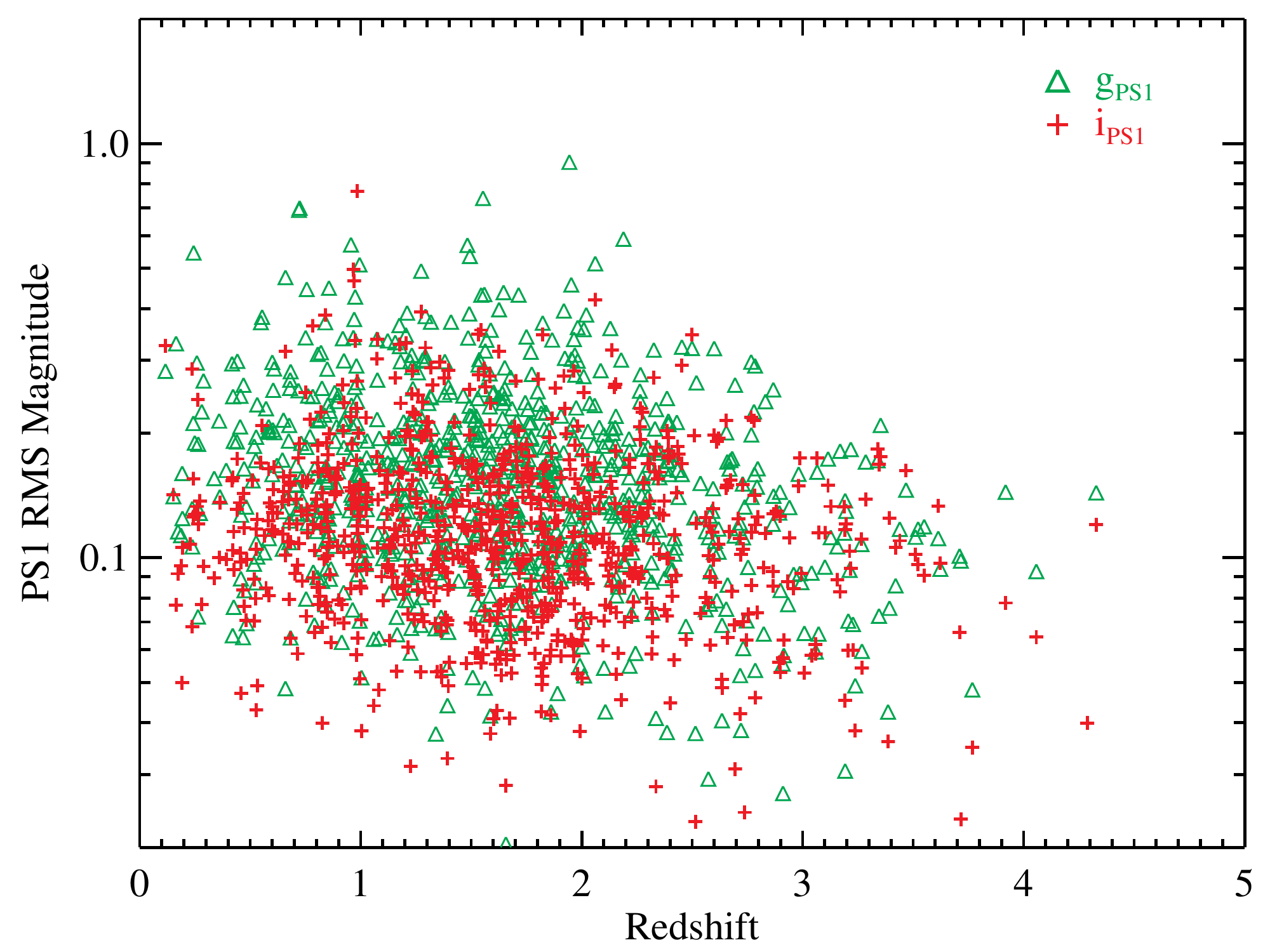}
     \caption{Intrinsic RMS magnitudes computed from the PS1 light curves during 2010-2013. We have corrected for photometric errors following the approach of \citet{Sesar_etal_2007}. These RMS magnitudes are typical of quasar continuum variability \citep[e.g.,][]{Sesar_etal_2007,MacLeod_etal_2010}.}
    \label{fig:ps1_rms}
\end{figure}

\begin{figure}
\centering
    \includegraphics[height=0.48\textwidth,angle=-90]{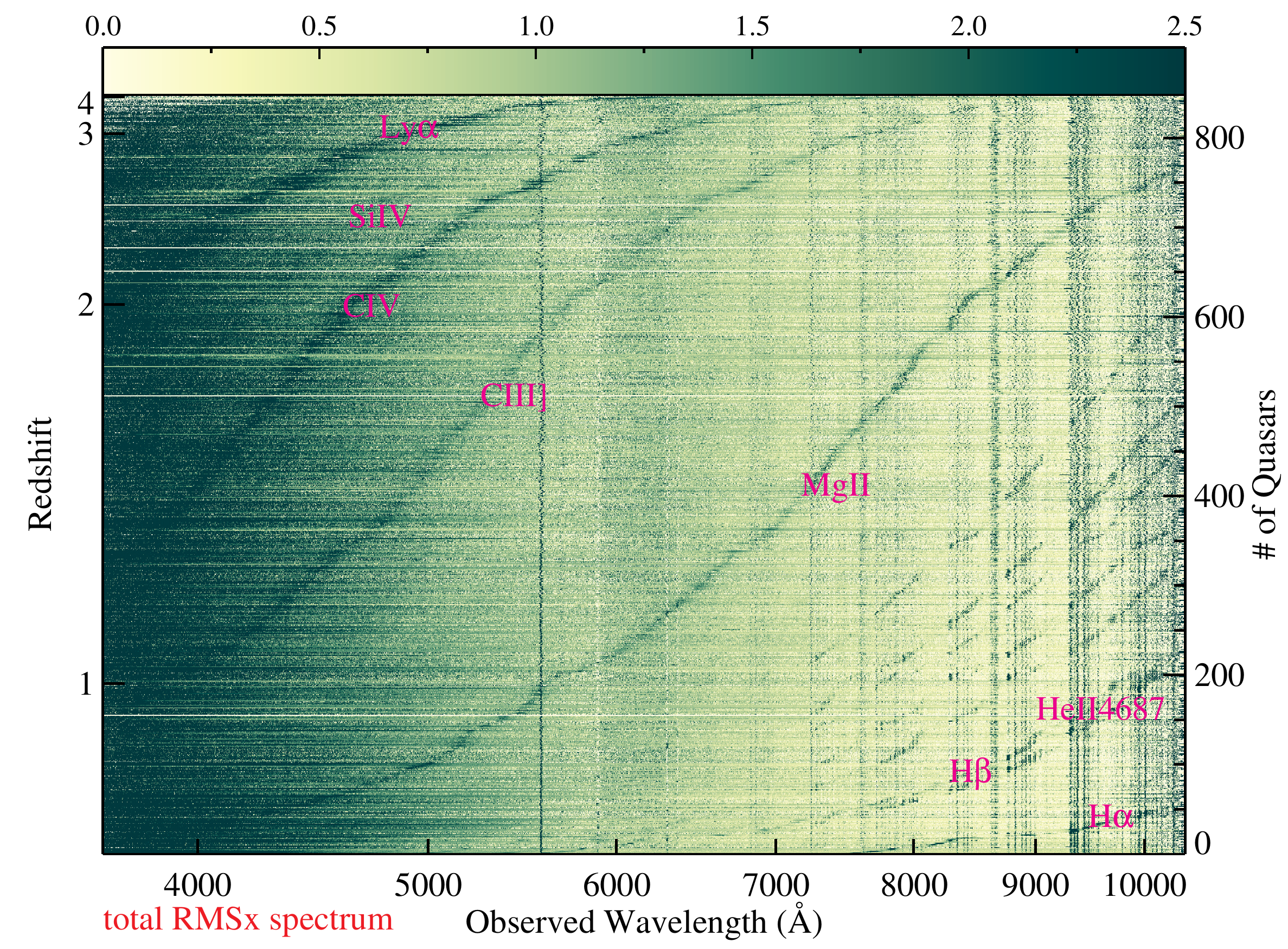}
     \caption{Intrinsic RMS spectra based on 2014 data for the 849 SDSS-RM quasars output by PrepSpec and ordered by redshift. Individual RMS spectra have been normalized to have a median value of 1. Variability in major broad emission lines is visible as trails (locally enhanced RMS variability) running from lower-left to upper-right. {Residuals around some narrow emission lines are also visible due to imperfect PrepSpec fits, which may be reduced in future PrepSpec versions.} Sky line residuals are manifested as vertical lines and become more severe near the red edge of the spectra. }
    \label{fig:spec_rms}
\end{figure}

The primary goal of SDSS-RM is to detect lags between the continuum and broad-line flux. Therefore the optical variability characteristics of our sample are of critical importance. We compute the intrinsic RMS magnitudes of SDSS-RM quasars using the early photometric light curves from PS1 during 2010--2013 in five bands \citep[$grizY_{\rm PS1}$;][]{Tonry_etal_2012a} for the MD07 field. Details of photometric calibration are described in \citet{Schlafly_etal_2012} and \citet{Magnier_etal_2013}. The intrinsic RMS magnitudes were computed following the approach in \citet[][eqn. 6]{Sesar_etal_2007} and subtracting the contribution from photometric errors. These (intrinsic) RMS magnitudes mostly measure the continuum variability level for our sample over a timescale of 4 years in the observed frame. Fig.\ \ref{fig:ps1_rms} shows the distribution of SDSS-RM quasars in the redshift-RMS mag plane (for $g_{\rm PS1}$ and $i_{\rm PS1}$; both are PSF magnitudes). {The median RMS magnitude for SDSS-RM quasars is 0.15 mag in $g_{\rm PS1}$ and 0.11 mag in $i_{\rm PS1}$}, typical of normal quasars \citep[e.g.,][]{Sesar_etal_2007,MacLeod_etal_2010}.

We further measure the spectral variability levels using the first-year spectroscopic observations. As detailed in \citet{Shen_etal_2016a}, PrepSpec fits a model that includes intrinsic variations in the continuum and broad emission lines to the time-resolved spectra, and outputs the light curves for continuum and broad-line emission. It also outputs the RMS spectrum computed from all spectroscopic epochs. During the PrepSpec fits, additional improvement was made to the flux calibration using available narrow line fluxes (assumed to be constant during the monitoring period) in individual epochs. 

We use the RMS spectra generated by PrepSpec on the 2014 SDSS-RM spectroscopy to construct a map of the spectral variability of SDSS-RM quasars in wavelength and redshift space in Fig.\ \ref{fig:spec_rms}. These individual RMS spectra were the measurement-error-corrected, maximum-likelihood estimate of the excess variability at each wavelength pixel (see below), and were referred to as the ``RMSx'' spectra in \citet{Shen_etal_2016a}. To make this map we normalize individual RMS spectra to have a median value of 1. It is apparent from this map that SDSS-RM quasars show significant variability in their broad emission lines, which is a necessary condition to measure a broad-line time lag with respect to the continuum. 

\begin{figure}
\centering
    \includegraphics[width=0.48\textwidth]{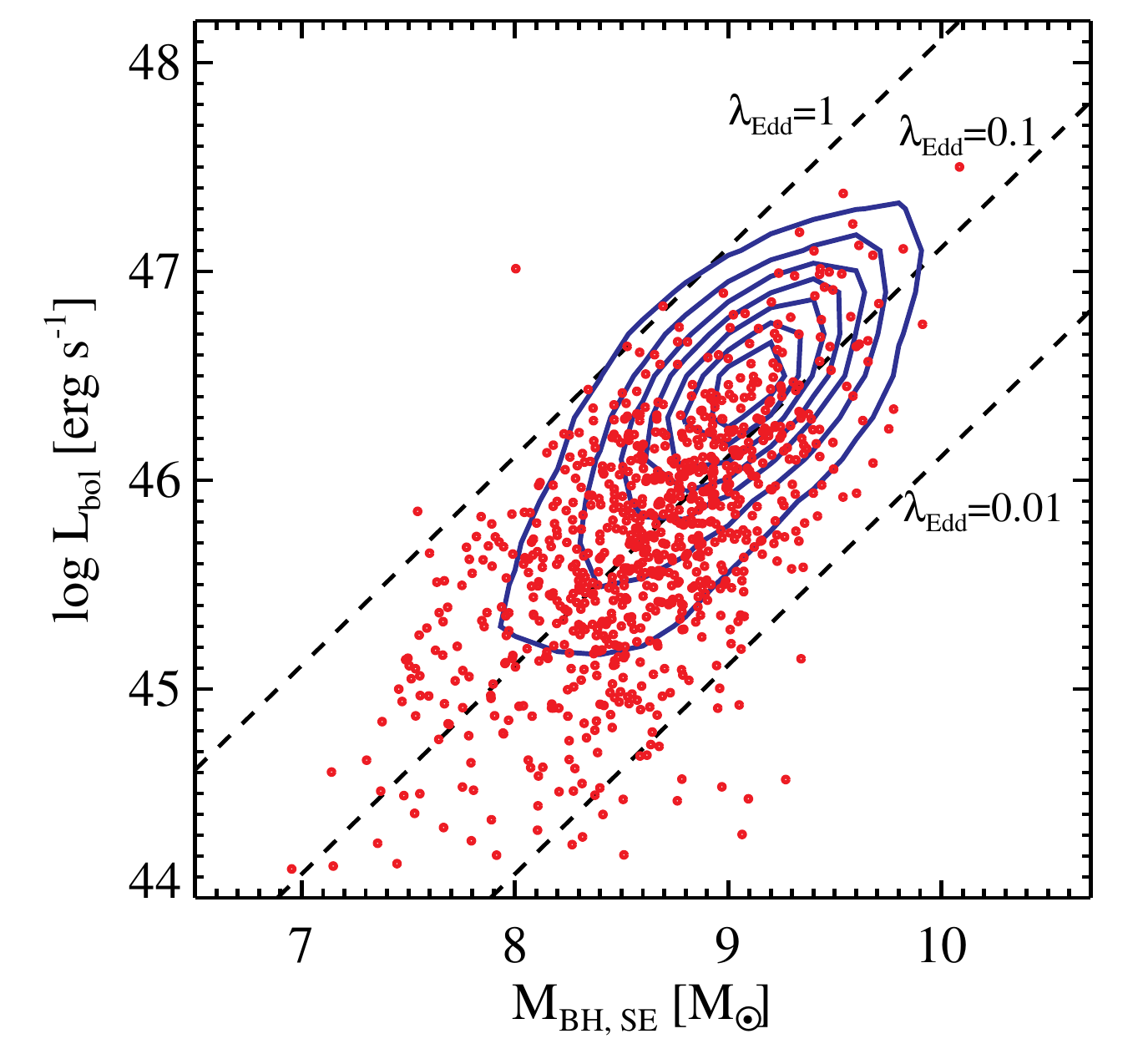}
     \caption{The BH mass and luminosity plane. The SDSS-RM sample is shown in red points. The blue contours are for the SDSS-DR7 quasars compiled in \citet{Shen_etal_2011}. Both samples use the single-epoch virial BH masses. The SDSS-RM quasars are on average $\sim 2$ mag fainter than the SDSS-DR7 sample, and have smaller Eddington ratios by 0.27 dex on average. The outlier with an extreme high Eddington ratio (RMID$=785$) is caused by a bad fit to a BAL quasar which largely underestimated the \CIV\ width and hence the single-epoch BH mass.}
    \label{fig:ml}
\end{figure}

Based on the PrepSpec light curves, we quantify the absolute and fractional (wrt to the average) continuum and broad-line variability for the SDSS-RM sample. We use a maximum-likelihood estimator to measure the excess variance of the light curve, as detailed below.

For a given time series $X_i$ with measurement error $\sigma_i$ and an unknown excess variance $\sigma_0^2$ resulting from intrinsic variability, we have
\begin{equation}
{\rm Var}[X_i] = \sigma_0^2 + \sigma_i^2=\frac{\sigma_0^2}{g_i},
\end{equation}
where 
\begin{equation}
g_i\equiv \frac{\sigma_0^2}{\sigma_0^2 + \sigma_i^2}=\frac{1}{1+(\sigma_i/\sigma_0)^2}
\end{equation}
quantifies the ``goodness'' of $X_i$ for measuring $\sigma_0^2$. $g_i$ varies from 0 for points with $\sigma_i\gg \sigma_0$ to 1 for points with $\sigma_i\ll \sigma_0$. The sum of $g_i$ over all data points then provides a ``goodness'' of measuring the intrinsic variability using the time series and approaches the total number of data points in the limit of $\sigma_i\ll \sigma_0$. We denote the sum of $g_i$ as ${\rm N\_RMS\_GOOD}$ as in Table \ref{tab:format}, which has an upper limit of 32 (i.e., total number of spectroscopic epochs in 2014). {The distribution of ${\rm N\_RMS\_GOOD}$ for our sample varies among different quantities. For continuum light curves, ${\rm N\_RMS\_GOOD}$ is mostly around 31. For emission lines, the median value of ${\rm N\_RMS\_GOOD}$ is lower and the dispersion is larger. For example, for broad \hbeta, the median ${\rm N\_RMS\_GOOD}$ is around 23 and the $16-84\%$ range is ${\rm N\_RMS\_GOOD}=13-29$. For the weaker \HeII\,4687 broad line, the median ${\rm N\_RMS\_GOOD}$ is around 21 and the $16-84\%$ range is ${\rm N\_RMS\_GOOD}=9-27$. }

{The likelihood function given $X_i$ and a constant model of $\mu\equiv \bracket{X_i}$} with both measurement errors and intrinsic variance is:
\begin{equation}
-2\ln L = \sum_{i=1}^{N}\frac{(X_i - \mu)^2}{\sigma_0^2+\sigma_i^2}+\sum_{i=1}^{N}\ln(\sigma_0^2+\sigma_i^2)\ .
\end{equation}

Minimizing the likelihood function and we obtain an estimate of $\sigma_0$ as
\begin{eqnarray}\label{eqn:sigma0}
\hat{\sigma_0}^2&=&\frac{\sum (X_i-\mu)^2g_i^2}{\sum g_i}\ ,\nonumber \\
{\rm Var}[\sigma_0^2]&=& \frac{\hat{\sigma_0}^4}{\sum g_i\frac{\displaystyle \sum (X_i-\mu)^2g_i^3}{\displaystyle \sum(X_i-\mu)^2g_i^2} - \sum g_i^2/2}\ .
\end{eqnarray}
To estimate the value of $\mu$, we use the optimal weights of individual data points based on $\sigma_i$ and $\sigma_0$:
\begin{equation}\label{eqn:mu}
\hat{\mu}=\frac{\displaystyle\sum\frac{X_i}{\sigma_0^2+\sigma_i^2}}{\displaystyle\sum\frac{1}{\sigma_0^2+\sigma_i^2}}=\frac{\sum X_ig_i}{\sum g_i}\ , {\rm Var}[\mu]=\frac{\sigma_0}{\sum g_i}\ .
\end{equation}
Equations (\ref{eqn:sigma0}) and (\ref{eqn:mu}) are solved iteratively. In Table \ref{tab:format} we denote $\sigma_0$ as RMS\_ML and its uncertainty as RMS\_ML\_ERR. One caveat is that the estimates for Var[$\mu$] and Var[$\sigma_0^2$] in Eqns (\ref{eqn:sigma0}) and (\ref{eqn:mu}) neglect the (usually small) covariance between $\hat{\mu}$ and $\hat{\sigma}_0^2$. 

To quantify how well the intrinsic variability is measured, we employ two separate metrics. The first metric is the SNR of the measured intrinsic RMS (SNR\_RMS\_ML), defined as RMS\_ML/RMS\_ML\_ERR. This metric may be better at low SNR, but it saturates,
near $\sqrt{2(N-1)}=7.9$ for $N=32$, when SNR is large, since there are only $N$ data points available to estimate the intrinsic variance. Therefore we define a second metric as ${\rm SNR2}=\sqrt{\chi^2-{\rm DOF}}$, where $\chi^2$ is relative to the optimal average using measurement errors ($\sigma_i$) only and ${\rm DOF}$ is the degree of freedom ($N-1$). Large intrinsic variability (wrt to measurement errors) will tend to produce a large SNR2. While there is no rigorous cut on SNR2 for robust detection of intrinsic variability, we recommend a threshold value of ${\rm SNR2}>20$ for a confident detection. 

In the main catalog described in \S\ref{sec:main_cat} and Table \ref{tab:format} we compile the intrinsic RMS variability for continuum flux at several wavelengths and for major broad lines. We also compile the fractional RMS variability relative to the average flux of the continuum and broad lines. However, we caution that the intrinsic variability may be overestimated if the feature is near the red edge of the spectral coverage, where significant sky line residuals are difficult to remove (see Fig.\ \ref{fig:spec_rms}).

Overall, the intrinsic fractional RMS spectral variability for the continuum and broad lines is at the $\sim 10\%$ level during the 2014 monitoring. Given the $\sim 5\%$ systematic uncertainty in our flux calibration \citep[][]{Shen_etal_2015a}, many quasars in our sample have well detected spectral variability. 



%

\section{Catalog Format}\label{sec:cat}

\subsection{Main Catalog}\label{sec:main_cat}

Table \ref{tab:format} provides the details of the compiled quasar properties and the catalog format. The associated data are provided in an online fits file. For each measured line we report the peak wavelength, FWHM, $\log L_{\rm line}$, rest-frame EW, and the line centroid computed from pixels above 50\% of the peak flux. The peak and centroid wavelengths ($\lambda_0^\prime$) were measured in the ``rest-frame'' using the pipeline redshift ZPIP, which can be easily converted to the rest-frame based on the improved redshift ZSYS, i.e., $\lambda_0 = \lambda_0^\prime(1+{\rm ZPIP})/(1+{\rm ZSYS})$, where $\lambda_0$ is the line wavelength in the rest-frame defined by ZSYS. 

We compile single-epoch virial BH masses based on the \hbeta, \MgII\ and \CIV\ lines, following fiducial recipes used in \citet{Shen_etal_2011}, which use the \citet{Vestergaard_Peterson_2006} calibrations for \hbeta\ and \CIV\ and our own calibration in that paper for \MgII. It is straightforward to use the compiled line widths and continuum luminosities to derive virial BH masses based on any other single-epoch estimators. 

For the fiducial single-epoch virial BH masses we adopt the estimates in the preference order of \hbeta, \MgII\ and \CIV. We refer the reader to the comprehensive review of \citet{Shen_2013} on the caveats of these different mass estimators. Bolometric luminosities are calculated from the $5100$\AA, $3000$\AA, and $1350$\AA\ monochromatic luminosities using bolometric corrections of 9.26, 5.15 and 3.81, respectively \citep[][]{Richards_etal_2006b,Shen_etal_2011}. We preferentially use the $3000$\AA\ luminosity to estimate the bolometric luminosity because it has less host contamination than the 5100\,\AA\ luminosity and suffers less from reddening and variability than the 1350\,\AA\ luminosity. We calculate Eddington ratios using the fiducial BH masses. These BH masses and Eddington ratios will be updated in the future from direct RM measurements. 

Using the fiducial single-epoch virial BH masses, Fig.\ \ref{fig:ml} compares the distribution of SDSS-RM quasars in the mass-luminosity plane with that of the SDSS-DR7 quasar sample in \citet{Shen_etal_2011}. The SDSS-RM quasars are $\sim 2$ mag fainter than SDSS-DR7 quasars, and they probe slightly lower Eddington ratios (with a median Eddington ratio of $\sim 0.1$).

\begin{longtable*}{llll}
\caption[notes]{FITS Catalog Format}\label{tab:format}\\
\hline \hline \\[0.2ex]
   \multicolumn{1}{l}{\textbf{Column}} &
   \multicolumn{1}{l}{\textbf{Format}} &
   \multicolumn{1}{l}{\textbf{Units}} &
   \multicolumn{1}{l}{\textbf{Description}} \\[0.2ex] \hline
   \\[0.2ex]
\endfirsthead
\multicolumn{4}{l}{{\tablename} \thetable{} -- Continued} \\[0.5ex]
  \hline \hline \\[0.2ex]
  \multicolumn{1}{l}{\textbf{Column}} &
  \multicolumn{1}{l}{\textbf{Format}} &
  \multicolumn{1}{l}{\textbf{Units}} &
  \multicolumn{1}{l}{\textbf{Description}} \\[0.2ex] \hline
  \\[0.2ex]
\endhead
  \hline
  \multicolumn{4}{l}{{Continued on Next Page\ldots}} \\
\endfoot
  \\[0.2ex] \hline \hline
\endlastfoot
RMID	&	LONG		&			& Object ID of SDSS-RM quasars \\                              
RA		&	DOUBLE	&	degree	& J2000 R.A. \\                                                
DEC		& DOUBLE 	& degree	& J2000 Decl.  \\                                              
ZPIP	& DOUBLE	& 		& pipeline redshift \\                                         
ZSYS	& DOUBLE	&			& Improved systemic redshift \\                                         
ZSYS\_ERR & DOUBLE & 	& Uncertainty in systemic redshift (systematic and statistical combined) \\                              
PSFMAG & DOUBLE[5] & mag  & SDSS PSF magnitudes in $(ugriz)_{\rm SDSS}$ ; undereddened for Galactic extinction \\   
MI   & DOUBLE   &  mag & Absolute $i$-band magnitude; undereddened for Galactic extinction \\                                     
MI\_Z2   & DOUBLE   & mag & $K$-corrected absolute $i$-band magnitude (normalized at $z=2$) \\                                     
BOSS\_TARGET1 & LONG64 &  & SDSS-III BOSS target bits \\                                       
BOSS\_TARGET2 & LONG64 &  & SDSS-III BOSS target bits \\                                       
ANCILLARY\_TARGET1 & LONG64 &  & SDSS-III BOSS ancillary target bits \\                        
ANCILLARY\_TARGET2 & LONG64 & & SDSS-III BOSS ancillary target bits \\                        
PRIMTARGET   & LONG64      & & SDSS-I/II DR7 primary target bits \\                       
SECTAEGET    & LONG64      &  & SDSS-I/II DR7 secondary target bits \\                     
OTHER\_TARGET  & STRING      &  & Other target flags \\                                  
PLATE\_ALL	& STRING         &  & Plate list of previous SDSS spectra (before SDSS-RM) \\          
FIBER\_ALL & STRING        &  & Fiber list of previous SDSS spectra (before SDSS-RM) \\           
MJD\_ALL   & STRING        &  & MJD list of previous SDSS spectra (before SDSS-RM) \\                    
PS1\_NMAG\_OK     & LONG[5]      &   & Number of good PS1 epochs in $(grizY)_{\rm PS1}$  \\                                              
PS1\_RMS\_MAG     & DOUBLE[5]    & mag  & intrinsic RMS magnitude of the PS1 light curves in $(grizY)_{\rm PS1}$  (PSF mag) \\                                              
ALLWISE1234     & DOUBLE[4]    &  mag &  WISE magnitudes from the ALLWISE release \\                                                
ALLWISE1234\_ERR & DOUBLE[4]    &  mag & Uncertainties in ALLWISE magnitudes  \\                                               
ALLWISE\_OFFSET  & DOUBLE       &  arcsec & Angular separation between SDSS and ALLWISE matches  \\                                               
UNWISE1234      & DOUBLE[4]    &   mag & Forced unWISE photometry at the SDSS position from \citet{Lang_etal_2016}  \\                                                
UNWISE1234\_ERR  & DOUBLE[4]    &  mag & Uncertainties in unWISE magnitudes  \\                                               
BAL\_FLAG        & LONG         &   & Broad absorption line flag; \\
                      &                     &   & 0=nonBAL or no coverage; 1=HiBAL; 2=LoBAL; 3=FeLoBAL  \\                                         
NABS            & LONG         &   &  Number of narrow absorption line systems \\                                                
ZABS            & DOUBLE[10]   &   &  Absorber redshifts of narrow absorption line systems (maximum 10) \\                                                
FIRST\_FR\_TYPE   & LONG         &   & FIRST radio morphology type; $-1$=not in FIRST footprint; \\
                      &                   &  & 0=FIRST undetected; 1=core-dominant; 2=lobe-dominant  \\                                              
FIRST\_FINT\_MJY  & DOUBLE       & mJy  &  FIRST integrated flux density at 20\,cm \\                                              
FINT\_REST6CM\_MJY\_OBS & DOUBLE  & mJy  &   Observed radio flux density at rest-frame 6\,cm \\                                             
LOGFNU2500A\_ERGS\_OBS & DOUBLE  &  \ergs${\rm cm}^{-2}{\rm Hz}^{-1}$ &  Observed optical flux density at rest-frame 2500\,\AA \\                                              
R\_6CM\_2500A    & DOUBLE        &   & Radio loudness $R\equiv f_{\nu,6{\rm cm}}/f_{\nu,2500A}$  \\ 
F\_H\_5100  & DOUBLE &  &   Host fraction at rest-frame 5100\,\AA\ from \citet{Shen_etal_2015b} \\
SIGMA  & DOUBLE &  $\kms$ & Host stellar velocity dispersion from \citet{Shen_etal_2015b} \\
SIGMA\_ERR & DOUBLE & $\kms$ & Uncertainty in SIGMA \\
SIGMA\_ERR\_WARNING & LONG & & 1 if SIGMA\_ERR may underestimate the systematic uncertainty  \\                                         
CONTI\_FIT      & DOUBLE[14]    &   &  Best-fit parameters for the continuum model \\                                               
CONTI\_FIT\_ERR  & DOUBLE[14]    &   & Uncertainties in the best-fit continuum parameters  \\         
CONTI\_REDCHI2  & DOUBLE        &   & Reduced $\chi^2$ for the continuum fit  \\                                                                                    
FEII\_UV        & DOUBLE[3]     &   &  Best-fit parameters for the UV \FeII\ model \\                                               
FEII\_UV\_ERR    & DOUBLE[3]     &  & Uncertainties in the best-fit UV \FeII\ model  \\                                              
FEII\_OPT       & DOUBLE[3]     &   &  Best-fit parameters for the optical \FeII\ model \\                                               
FEII\_OPT\_ERR   & DOUBLE[3]     &   & Uncertainties in the best-fit optical \FeII\ model  \\                                              
LOGL1350       & DOUBLE        &  [\ergs] & Continuum luminosity at rest-frame 1350\,\AA  \\                                                
LOGL1350\_ERR   & DOUBLE        & [\ergs]  &  Uncertainty in LOGL1350  \\                                               
LOGL1700       & DOUBLE        & [\ergs]  &  Continuum luminosity at rest-frame 1700\,\AA \\                                                
LOGL1700\_ERR   & DOUBLE        & [\ergs]  & Uncertainty in LOGL1700  \\                                               
LOGL3000       & DOUBLE        & [\ergs]  & Continuum luminosity at rest-frame 3000\,\AA  \\                                                
LOGL3000\_ERR   & DOUBLE        & [\ergs]  &  Uncertainty in LOGL3000 \\                                               
LOGL5100       & DOUBLE        & [\ergs]  & Continuum luminosity at rest-frame 5100\,\AA  \\                                                
LOGL5100\_ERR   & DOUBLE        & [\ergs]  &  Uncertainty in LOGL5100 \\ 
LOGLBOL       & DOUBLE        & [\ergs]  & Bolometric luminosity  \\                                                
LOGLBOL\_ERR   & DOUBLE        & [\ergs]  &  Uncertainty in LOGLBOL \\                                              
REW\_FE\_4434\_4684 &  DOUBLE    & \AA    &  Rest-frame equivalent width of optical \FeII\ within 4434-4684\,\AA \\                                              
REW\_FE\_4434\_4684\_ERR & DOUBLE  & \AA & Uncertainty in  REW\_FE\_4434\_4684 \\  
REW\_FE\_2250\_2650 &  DOUBLE    & \AA    &  Rest-frame equivalent width of UV \FeII\ within 2250-2650\,\AA \\                                              
REW\_FE\_2250\_2650\_ERR & DOUBLE  & \AA & Uncertainty in  REW\_FE\_2250\_2650 \\                                                 
SII6718         & DOUBLE[5]  & \AA, $\kms$, [\ergs], \AA, \AA  & peak wavelength, FWHM, $\log L_{\rm line}$, rest-frame EW, top 50\% flux centroid \\                                                   
HALPHA          & DOUBLE[5]  & ...  & For the entire \halpha\ profile (narrow and broad lines combined)  \\                                                   
HALPHA\_BR       & DOUBLE[5]  & ...  & For the broad \halpha\ component \\                    
NII6585          & DOUBLE[5]  & ...  & For the narrow \NIIb\ component \\                              
HBETA           & DOUBLE[5]  &  ... &  For the entire \hbeta\ profile (narrow and broad lines combined) \\                                                   
HBETA\_BR        & DOUBLE[5]  & ...  & For the broad \hbeta\ component \\                                                  
HEII4687        & DOUBLE[5]  &  ... & For the entire \HeIIopt\ profile (narrow and broad lines combined) \\                                                   
HEII4687\_BR     & DOUBLE[5]  & ...  & For the broad \HeIIopt\ component \\                                                  
OIII5007        & DOUBLE[5]  & ...  & For the entire \OIIIb\ profile \\                                                   
OIII5007C       & DOUBLE[5]  & ...  & For the core \OIIIb\ profile  \\                                                   
CAII3934        & DOUBLE[5]  & ...  &  For the \CaII\ K absorption line \\                                                   
OII3728         & DOUBLE[5]  &  ... & ... \\                                                   
NEV3426         & DOUBLE[5]  & ...  & ... \\                                                   
MGII            & DOUBLE[5]  & ...  & For the entire \MgII\ profile (narrow and broad lines combined) \\                                                   
MGII\_BR         & DOUBLE[5]  & ...  & For the broad \MgII\ component \\                                                  
CIII\_ALL        & DOUBLE[5]  & ...  & For the entire \CIII\ complex (\CIII, \SiIII, \AlIII)  \\                                                  
CIII\_BR         & DOUBLE[5]  & ...   & For the broad \CIII\ component \\                                                  
SIIII1892       & DOUBLE[5]  &  ... & ... \\                                                   
ALIII1857       & DOUBLE[5]  & ...  & ... \\                                                   
NIII1750        & DOUBLE[5]  & ...  & ...  \\                                                   
CIV             & DOUBLE[5]  & ...   & ...  \\                                                   
HEII1640        & DOUBLE[5]  & ...  & For the entire \HeIIuv\ profile (narrow and broad lines combined) \\                                                   
HEII1640\_BR     & DOUBLE[5]  & ...  & For the broad \HeIIuv\ component \\                                                  
SIIV\_OIV        & DOUBLE[5]  & ...  &  For the 1400\,\AA\ complex \\                                                  
OI1304          & DOUBLE[5]  &  ... & ... \\                                                   
LYA             & DOUBLE[5]  &  ... & ... \\                                                   
NV1240          & DOUBLE[5]  & ...  &  ... \\                                                   
SII6718\_ERR     & DOUBLE[5]  & \AA, $\kms$, [\ergs], \AA, \AA  & Measurement errors in \SIIa \\                                                  
HALPHA\_ERR      & DOUBLE[5]  & ...  & ...  \\                                                  
HALPHA\_BR\_ERR   & DOUBLE[5]  & ...  &  ... \\  
NII6585\_ERR     & DOUBLE[5]  & ... & ... \\                                               
HBETA\_ERR       & DOUBLE[5]  & ...  & ...  \\                                                  
HBETA\_BR\_ERR    & DOUBLE[5]  & ...  & ...  \\                                                   
HEII4687\_ERR    & DOUBLE[5]  & ...  & ... \\                                                  
HEII4687\_BR\_ERR & DOUBLE[5]  & ...  & ...  \\                                                 
OIII5007\_ERR    & DOUBLE[5]  & ...  & ... \\                                                  
OIII5007C\_ERR   & DOUBLE[5]  & ...  & ... \\                                                   
CAII3934\_ERR    & DOUBLE[5]  & ...  & ... \\                                                  
OII3728\_ERR     & DOUBLE[5]  & ...  &  ... \\                                                  
NEV3426\_ERR     & DOUBLE[5]  & ...  & ... \\                                                  
MGII\_ERR        & DOUBLE[5]  & ...  & ... \\                                                  
MGII\_BR\_ERR     & DOUBLE[5]  & ...  & ... \\                                                 
CIII\_ALL\_ERR    & DOUBLE[5]  & ...  & ... \\                                                 
CIII\_BR\_ERR     & DOUBLE[5]  & ...  & ... \\                                                 
SIIII1892\_ERR   & DOUBLE[5]  &  ... & ... \\                                                  
ALIII1857\_ERR   & DOUBLE[5]  & ...  & ... \\                                                  
NIII1750\_ERR    & DOUBLE[5]  & ...  & ... \\                                                  
CIV\_ERR         & DOUBLE[5]  & ...  & ... \\                                                  
HEII1640\_ERR    & DOUBLE[5]  & ...  & ... \\                                                  
HEII1640\_BR\_ERR & DOUBLE[5]  &  ... & ... \\                                                 
SIIV\_OIV\_ERR    & DOUBLE[5]  & ...  & ... \\                                                 
OI1304\_ERR      & DOUBLE[5]  &  ... & ... \\                                                  
LYA\_ERR         & DOUBLE[5]  & ...   & ... \\                                                  
NV1240\_ERR      & DOUBLE[5]  & ...   & ... \\                                                  
LOGBH\_CIV\_VP06  & DOUBLE     &  [$M_\odot$] & Single-epoch BH mass based on \CIV\ \citep{Vestergaard_Peterson_2006} \\                                                 
LOGBH\_CIV\_VP06\_ERR & DOUBLE  & ...  & ... \\                                                 
LOGBH\_MGII\_S11  & DOUBLE     & ...  & Single-epoch BH mass based on \MgII\ \citep{Shen_etal_2011} \\                                                 
LOGBH\_MGII\_S11\_ERR & DOUBLE  & ...  & ... \\                                                 
LOGBH\_HB\_VP06  & DOUBLE      & ...  & Single-epoch BH mass based on \hbeta\ \citep{Vestergaard_Peterson_2006} \\                                                 
LOGBH\_HB\_VP06\_ERR & DOUBLE   & ... & ... \\         
LOGBH   & DOUBLE      & ...  & Fiducial single-epoch BH mass \\                                                 
LOGBH\_ERR & DOUBLE   & ... &  \\                                          
LOGEDD\_RATIO & DOUBLE & ... & Eddington ratio based on fiducial single-epoch BH mass \\
LOGEDD\_RATIO\_ERR & DOUBLE & & \\ 
RMS\_ML\_C1700 & DOUBLE & ${\rm erg\,s^{-1}\,cm^{-2}\textrm{\AA}^{-1}}$ & (Maximum-likelihood) RMS variability for rest 1700\,\AA\ continuum    \\
RMS\_ML\_C1700\_ERR & DOUBLE & ${\rm erg\,s^{-1}\,cm^{-2}\textrm{\AA}^{-1}}$ &  \\
SNR\_RMS\_ML\_C1700 &  DOUBLE &  & RMS\_ML\_C1700/RMS\_ML\_C1700\_ERR \\
RMS\_ML\_FRAC\_C1700 & DOUBLE &  & Fractional (wrt average) RMS variability for rest 1700\,\AA\ continuum \\
RMS\_ML\_FRAC\_C1700\_ERR & DOUBLE & & \\
N\_RMS\_GOOD\_C1700 & FLOAT & & Effective number of good data points in estimating RMS\_ML\_C1700  \\
SNR2\_C1700 & DOUBLE & & Alternative measure of the variability for rest 1700\,\AA\ continuum \\
RMS\_ML\_C3000 & DOUBLE & ${\rm erg\,s^{-1}\,cm^{-2}\textrm{\AA}^{-1}}$ & (Maximum-likelihood) RMS variability for rest 3000\,\AA\ continuum    \\
RMS\_ML\_C3000\_ERR & DOUBLE & ${\rm erg\,s^{-1}\,cm^{-2}\textrm{\AA}^{-1}}$ &  \\
SNR\_RMS\_ML\_C3000 &  DOUBLE &  & RMS\_ML\_C3000/RMS\_ML\_C3000\_ERR \\
RMS\_ML\_FRAC\_C3000 & DOUBLE &  & Fractional (wrt average) RMS variability for rest 3000\,\AA\ continuum \\
RMS\_ML\_FRAC\_C3000\_ERR & DOUBLE & & \\
N\_RMS\_GOOD\_C3000 & FLOAT & & Effective number of good data points in estimating RMS\_ML\_C3000  \\
SNR2\_C3000 & DOUBLE & & Alternative measure of the variability for rest 3000\,\AA\ continuum \\
RMS\_ML\_C5100 & DOUBLE & ${\rm erg\,s^{-1}\,cm^{-2}\textrm{\AA}^{-1}}$ & (Maximum-likelihood) RMS variability for rest 5100\,\AA\ continuum    \\
RMS\_ML\_C5100\_ERR & DOUBLE & ${\rm erg\,s^{-1}\,cm^{-2}\textrm{\AA}^{-1}}$ &  \\
SNR\_RMS\_ML\_C5100 &  DOUBLE &  & RMS\_ML\_C5100/RMS\_ML\_C5100\_ERR \\
RMS\_ML\_FRAC\_C5100 & DOUBLE &  & Fractional (wrt average) RMS variability for rest 5100\,\AA\ continuum \\
RMS\_ML\_FRAC\_C5100\_ERR & DOUBLE & & \\
N\_RMS\_GOOD\_C5100 & FLOAT & & Effective number of good data points in estimating RMS\_ML\_C5100  \\
SNR2\_C5100 & DOUBLE & & Alternative measure of the variability for rest 5100\,\AA\ continuum \\
RMS\_ML\_HA & DOUBLE & ${\rm erg\,s^{-1}\,cm^{-2}}$ & (Maximum-likelihood) RMS variability for broad \halpha    \\
RMS\_ML\_HA\_ERR & DOUBLE & ${\rm erg\,s^{-1}\,cm^{-2}}$ &  \\
SNR\_RMS\_ML\_HA &  DOUBLE &  & RMS\_ML\_HA/RMS\_ML\_HA\_ERR \\
RMS\_ML\_FRAC\_HA & DOUBLE &  & Fractional (wrt average) RMS variability for broad \halpha \\
RMS\_ML\_FRAC\_HA\_ERR & DOUBLE & & \\
N\_RMS\_GOOD\_HA & FLOAT & & Effective number of good data points in estimating RMS\_ML\_HA  \\
SNR2\_HA & DOUBLE & & Alternative measure of the variability for broad \halpha \\
RMS\_ML\_HB & DOUBLE & ${\rm erg\,s^{-1}\,cm^{-2}}$ & (Maximum-likelihood) RMS variability for broad \hbeta    \\
RMS\_ML\_HB\_ERR & DOUBLE & ${\rm erg\,s^{-1}\,cm^{-2}}$ &  \\
SNR\_RMS\_ML\_HB &  DOUBLE &  & RMS\_ML\_HB/RMS\_ML\_HB\_ERR \\
RMS\_ML\_FRAC\_HB & DOUBLE &  & Fractional (wrt average) RMS variability for broad \hbeta \\
RMS\_ML\_FRAC\_HB\_ERR & DOUBLE & & \\
N\_RMS\_GOOD\_HB & FLOAT & & Effective number of good data points in estimating RMS\_ML\_HB  \\
SNR2\_HB & DOUBLE & & Alternative measure of the variability for broad \HeII\,4687 \\
RMS\_ML\_HEII4687 & DOUBLE & ${\rm erg\,s^{-1}\,cm^{-2}}$ & (Maximum-likelihood) RMS variability for broad \HeII\,4687    \\
RMS\_ML\_HEII4687\_ERR & DOUBLE & ${\rm erg\,s^{-1}\,cm^{-2}}$ &  \\
SNR\_RMS\_ML\_HEII4687 &  DOUBLE &  & RMS\_ML\_HEII4687/RMS\_ML\_HEII4687\_ERR \\
RMS\_ML\_FRAC\_HEII4687 & DOUBLE &  & Fractional (wrt average) RMS variability for broad \HeII\,4687 \\
RMS\_ML\_FRAC\_HEII4687\_ERR & DOUBLE & & \\
N\_RMS\_GOOD\_HEII4687 & FLOAT & & Effective number of good data points in estimating RMS\_ML\_HEII4687  \\
SNR2\_HEII4687 & DOUBLE & & Alternative measure of the variability for broad \HeII\,4687 \\
RMS\_ML\_MGII & DOUBLE & ${\rm erg\,s^{-1}\,cm^{-2}}$ & (Maximum-likelihood) RMS variability for broad \MgII   \\
RMS\_ML\_MGII\_ERR & DOUBLE & ${\rm erg\,s^{-1}\,cm^{-2}}$ &  \\
SNR\_RMS\_ML\_MGII &  DOUBLE &  & RMS\_ML\_MGII/RMS\_ML\_MGII\_ERR \\
RMS\_ML\_FRAC\_MGII & DOUBLE &  & Fractional (wrt average) RMS variability for broad \MgII \\
RMS\_ML\_FRAC\_MGII\_ERR & DOUBLE & & \\
N\_RMS\_GOOD\_MGII & FLOAT & & Effective number of good data points in estimating RMS\_ML\_MGII  \\
SNR2\_MGII & DOUBLE & & Alternative measure of the variability for broad \MgII \\
RMS\_ML\_CIII & DOUBLE & ${\rm erg\,s^{-1}\,cm^{-2}}$ & (Maximum-likelihood) RMS variability for the full \CIII\ complex   \\
RMS\_ML\_CIII\_ERR & DOUBLE & ${\rm erg\,s^{-1}\,cm^{-2}}$ &  \\
SNR\_RMS\_ML\_CIII &  DOUBLE &  & RMS\_ML\_CIII/RMS\_ML\_CIII\_ERR \\
RMS\_ML\_FRAC\_CIII & DOUBLE &  & Fractional (wrt average) RMS variability for the full \CIII\ complex \\
RMS\_ML\_FRAC\_CIII\_ERR & DOUBLE & & \\
N\_RMS\_GOOD\_CIII & FLOAT & & Effective number of good data points in estimating RMS\_ML\_CIII  \\
SNR2\_CIII & DOUBLE & & Alternative measure of the variability for the full \CIII\ complex \\
RMS\_ML\_CIV & DOUBLE & ${\rm erg\,s^{-1}\,cm^{-2}}$ & (Maximum-likelihood) RMS variability for broad \CIV   \\
RMS\_ML\_CIV\_ERR & DOUBLE & ${\rm erg\,s^{-1}\,cm^{-2}}$ &  \\
SNR\_RMS\_ML\_CIV &  DOUBLE &  & RMS\_ML\_CIV/RMS\_ML\_CIV\_ERR \\
RMS\_ML\_FRAC\_CIV & DOUBLE &  & Fractional (wrt average) RMS variability for broad \CIV \\
RMS\_ML\_FRAC\_CIV\_ERR & DOUBLE & & \\
N\_RMS\_GOOD\_CIV & FLOAT & & Effective number of good data points in estimating RMS\_ML\_CIV  \\
SNR2\_CIV & DOUBLE & & Alternative measure of the variability for broad \CIV \\
RMS\_ML\_LYA & DOUBLE & ${\rm erg\,s^{-1}\,cm^{-2}}$ & (Maximum-likelihood) RMS variability for broad \lya   \\
RMS\_ML\_LYA\_ERR & DOUBLE & ${\rm erg\,s^{-1}\,cm^{-2}}$ &  \\
SNR\_RMS\_ML\_LYA &  DOUBLE &  & RMS\_ML\_LYA/RMS\_ML\_LYA\_ERR \\
RMS\_ML\_FRAC\_LYA & DOUBLE &  & Fractional (wrt average) RMS variability for broad \lya \\
RMS\_ML\_FRAC\_LYA\_ERR & DOUBLE & & \\
N\_RMS\_GOOD\_LYA & FLOAT & & Effective number of good data points in estimating RMS\_ML\_LYA  \\
SNR2\_LYA & DOUBLE & & Alternative measure of the variability for broad \lya \\
COMMENT        & STRING     &  ...  &  Comments on individual objects \\                                                  
\end{longtable*}
\tablecomments{\footnotesize (1) $K-$corrections are the same as in \citet{Richards_etal_2006a}; (2) Bolometric luminosities are computed using bolometric corrections of 9.26, 5.15, and 3.81
\citep{Richards_etal_2006b} using the $5100$\AA, $3000$\AA, and $1350$\AA\ monochromatic luminosities respectively, with $3000$\AA\ luminosity being the highest priority; (3) Uncertainties are measurement errors only; (4) Unless otherwise stated, unmeasurable values are indicated with zero for a quantity and $-1$ for its associated error, except for LOGEDD\_RATIO where the unmeasurable values are $-99$; (5) In general the FWHM of the total line profile is smaller than that of the broad-only component. However, in cases where the narrow and broad components are significantly offset in velocity (e.g., \HeII1640\ in some objects), the total line FWHM can be larger than the broad-line-only FWHM. }

\subsection{Supplemental Catalogs}

As described in previous sections, we have compiled additional properties for the SDSS-RM sample in several ancillary catalogs. Below are the notes on these supplemental catalogs. All data files and their documentation are provided along with this paper and archived on the SDSS-RM data server. 

\begin{enumerate}

\item[$\bullet$] {\it allqso\_sdssrm.fits} A FITS table of all 1214 known quasars in the 7\,deg$^2$ SDSS-RM field. Only 849 of them received a fiber in the SDSS-RM spectroscopy. This table lists the basic target information of these quasars. 

\item[$\bullet$] {\it QSObased\_Expanded\_SDSSRM\_107.fits} The narrow \MgII/\FeII\ absorber catalog for SDSS-RM quasars, following the methodology outlined in \citet{Zhu_Menard_2013}. Each entry corresponds to one quasar. The search for narrow absorbers includes systems that have absorber redshift close to the quasar systemic redshift ($|\Delta z|<0.04$). \MgII\ absorbers blueshifted from the quasar by $\Delta z>0.04$ and also redward of \CIV\ by $\Delta z>0.02$ are of high purity. \MgII\ absorbers with $|\Delta z|<0.04$ or those at wavelength blueward of \CIV, or those with \FeII\ detection but no \MgII\ detections (likely due to bad pixels), while included in this catalog, should be treated with caution, and may contain a small fraction of false positives (mainly \CIV\ absorbers). 

For convenience, we also provide a version of the absorber catalog organized by absorbers ({\it Expanded\_SDSSRM\_107.fits}), i.e., each entry corresponds to one absorber system. 

\item[$\bullet$] {\it rmqso32\_aegis\_multi\_lambda.fits} Multi-wavelength data compiled from \citet{Nandra_etal_2015} for 32 SDSS-RM quasars in the AGEIS field. 

\item[$\bullet$] {\it spitzer\_seip\_rm\_match\_1.5arcsec.fits} Spitzer IRAC and MIPS data from the Spitzer Enhanced Imaging Products (SEIP) source list for 176 SDSS-RM quasars, with a matching radius of 1\arcsec.5. This file also compiles infrared fluxes (if available) from 2MASS \citep{Skrutskie_etal_2006}. 

\item[$\bullet$] {\it spec\_2014\_BALrobust.csv} List of 95 BALQSOs (including mini-BALQSOs) identified from the first-year coadded spectroscopy. This file includes BAL flags on \CIV, \AlIII, \MgII, and \ion{Fe}{2}/\ion{Fe}{3}. It also includes notes on individual objects. 

\item[$\bullet$] {\it PS1\_MD07\_LC\_sdssrm.fits} PS1 Medium Deep light curves for the SDSS-RM quasars used to compute PS1\_NMAG\_OK and PS1\_RMS\_MAG in the main catalog. Note this is the unofficial release of the PS1 MD07 data, which was approved by the PS1 collaboration. These photometric light curves may differ slightly from the final official release of the PS1 Medium Deep field data. 

\end{enumerate}

%
%


\section{Summary}\label{sec:sum}

We have presented a detailed characterization of the spectral and optical variability properties of a representative quasar sample from the SDSS-RM project. The compiled main and supplemental catalogs will serve as the basis for future SDSS-RM work that study quasars and their host galaxies. 

The SDSS-RM sample probes a diverse range in quasar properties over a broad redshift range of $0.1<z<4.5$. The high-SNR coadded spectroscopy allowed robust measurements of the spectral properties, and the multi-epoch spectroscopic data provided important constraints on the spectral variability of our sample. 

Many SDSS-RM quasars have well-detected variability in their continuum and broad emission lines, which is a necessary (but not sufficient) condition for lag measurements. 


The SDSS-RM project will continue to carry out monitoring imaging and spectroscopy through 2020, as well as other dedicated multi-wavelength follow-up programs in the same field. These future data sets will provide additional information on the quasar sample, and will be included in future data releases of SDSS-RM.  

\acknowledgements


We thank the anonymous referee for comments that improved the manuscript. YS acknowledges support from an Alfred P. Sloan Research Fellowship and NSF grant AST-1715579. PH acknowledges support from the Natural Sciences and Engineering Research Council of Canada (NSERC), funding reference number 2017-05983. WNB acknowledges support from NSF grant AST-1516784. CJG, WNB, and DPS acknowledge support from NSF grant AST-1517113. Funding for SDSS-III has been provided by the Alfred P. Sloan Foundation, the Participating Institutions, the National Science Foundation, and the U.S.
Department of Energy Office of Science. The SDSS-III web site is http://www.sdss3.org/.

SDSS-III is managed by the Astrophysical Research Consortium for the
Participating Institutions of the SDSS-III Collaboration including the
University of Arizona, the Brazilian Participation Group, Brookhaven National
Laboratory, University of Cambridge, Carnegie Mellon University, University
of Florida, the French Participation Group, the German Participation Group,
Harvard University, the Instituto de Astrofisica de Canarias, the Michigan
State/Notre Dame/JINA Participation Group, Johns Hopkins University, Lawrence
Berkeley National Laboratory, Max Planck Institute for Astrophysics, Max
Planck Institute for Extraterrestrial Physics, New Mexico State University,
New York University, Ohio State University, Pennsylvania State University,
University of Portsmouth, Princeton University, the Spanish Participation
Group, University of Tokyo, University of Utah, Vanderbilt University,
University of Virginia, University of Washington, and Yale University.

\appendix

%

\section{A. Details of the spectral fitting code (QSOFIT)}\label{sec:app2}

We here provide a brief user's guide to the spectral fitting code ({\tt qsofit}; written by Y. Shen) used to measure spectral properties of SDSS-RM quasars, which is a general-purpose code for quasar spectral fits. The code is written in IDL; a python version of the code is developed and made public \citep[][]{pyqsofit}. The package includes the main routine, \FeII\ templates, an input line-fitting parameter file, and ancillary routines used to extract spectral measurements from the fits. Monte Carlo estimation of the measurement uncertainties of the fitting results can be conducted with the same fitting code. The software requires the installation of the {\it idlutils} package\footnote{http://www.sdss3.org/dr8/software/idlutils.php}, which includes the widely-used MPFIT IDL package \citep{Markwardt_2009} to perform the $\chi^2$ minimization.  

The code takes an input spectrum (observed-frame wavelength, flux density and error arrays) and the redshift as input parameters, performs the fitting in the rest-frame, and outputs the best-fit parameters and quality assessment (QA) plots to the paths specified by the user. The fitting results are stored as a binary table in an IDL structure format. The input flux density and errors are assumed to be in units of $10^{-17}\,{\rm erg\,s^{-1}cm^{-2}\textrm{\AA}^{-1}}$ per SDSS default. Since the fitting is performed in the restframe of the quasar, the model spectrum ($f_{\lambda}$) should be multiplied by $(1+z)$ when computing the monochromatic continuum luminosity $\lambda f_{\lambda}$ or the integrated line luminosity. 

The code uses an input line-fitting parameter file (qsoline*.par) to specify the fitting range and parameter constraints of the individual emission line components. An example of such a file is provided in the package. Within the code, the user can switch on/off components to fit to the pseudo-continuum. For example, for some objects the UV/optical \FeII\ emission cannot be well constrained and the user can exclude this component in the continuum fit. The code is highly flexible and can be modified to meet the specific needs of the user. For example, if a local fit around certain emission lines is required rather than a global fit, the user can truncate the input spectrum before feeding it to {\tt qsofit}. 

An example calling sequence of fitting to a quasar spectrum:

\texttt{IDL> qsofit, wave\_obs, flux, err, z, /psplot, /fits, emparfile='qsoline.par'}\\

The package also provides several sub-routines to compute the spectral quantities from the model fits:
\begin{enumerate}

\item[$\bullet$] reconstruct the pseudo-continuum flux (\FeII\ excluded):

\texttt{IDL> para = mrdfits('output.fits',1) \% read the fits structure output by qsofit\\
IDL> conti\_fit = para.conti\_fit \% parameter array for the pseudo-continuum fit\\
IDL> conti\_flux = f\_conti\_only(wave, conti\_fit[6:*]) \% the first 6 elements in contifit are reserved for the FeII model}

\item[$\bullet$] reconstruct the \FeII\ flux:

\texttt{IDL> f\_FeII\_uv = fe\_flux\_mgii(wave, conti\_fit[0:2])\\
IDL> f\_FeII\_opt = fe\_flux\_balmer(wave, conti\_fit[3:5])}

\item[$\bullet$] reconstruct the model line flux: 

\texttt{IDL> line\_fit = para.line\_fit \% parameter array for the emission line fit\\
IDL> linename = para.linename \% get the corresponding line names\\
IDL> ind = where(strmatch(linename, 'CIV')) \% indices for the CIV line\\
IDL> pp = line\_fit[ind]\\
IDL> line\_flux = manygauss(alog(wave), pp) \% model line flux}

\item[$\bullet$] obtain line properties from the multi-Gaussian fit:

\texttt{IDL> result = get\_multi\_gaussian\_prop(pp,/diet)\\
\% peak wavelength = exp(result[0])\\
\% FWHM = result[1]*3d5\\
\% line flux = result[2]*(1+z); in units of $10^{-17}{\rm erg\,s^{-1}cm^{-2}}$ \\
\% top 50\% flux centroid = exp(result[5])}

\end{enumerate}


\bibliography{/Users/yshen/Research/refs}

\end{document}